\documentclass[11pt,a4paper]{article}
\usepackage{url}
\usepackage{hyperref}
\usepackage{graphicx}  \usepackage{mathptmx}
\usepackage{amsmath, amssymb}
\usepackage{gensymb}
\usepackage{enumitem}
\usepackage{tikz}
\usepackage{tikz-cd}
\usepackage{wrapfig}
\usetikzlibrary{decorations.pathmorphing}
\tikzset{snake it/.style={decorate, decoration=snake}}
\usepackage{xpatch}
\topmargin = - 1.5 cm
\makeatletter
\@addtoreset{equation}{section}
\makeatother

\newcommand{\id}[1]{\ensuremath{\mathrm{id}}}

\newcommand{\third}{\mbox{\footnotesize $\frac{1}{3}$}}

\newcommand{\half}{\mbox{\footnotesize $\frac{1}{2}$}}
\newcommand{\hi}[1]{\emph{\textbf{#1}}}
\newcommand{\qm}{quantum mechanics}
\newcommand{\er}{\eqref}
 
\newcommand{\beq}{\begin{equation}}
\newcommand{\eeq}{\end{equation}} 
\newcommand{\bea}{\begin{eqnarray}}
\newcommand{\eea}{\end{eqnarray}}

\newcommand{\ul}{\underline}

\newcommand{\raw}{\rightarrow}

 \newcommand{\Raw}{\Rightarrow}

\newcommand{\x}{\times}

\newcommand{\gm}{\gamma} \newcommand{\Gm}{\Gamma}
\newcommand{\dl}{\delta} \newcommand{\Dl}{\Delta}

\newcommand{\io}{\iota} 
\newcommand{\lm}{\lambda} 
\newcommand{\rh}{\rho} \newcommand{\sg}{\sigma}
  
 \newcommand{\phv}{\varphi}
\newcommand{\ch}{\ch} \newcommand{\ps}{\psi} 
\newcommand{\om}{\omega} \newcommand{\Om}{\Omega}

 \newcommand{\CI}{{\mathcal I}}

 \newcommand{\R}{{\mathbb R}}
 
  \makeatletter
 \newcommand{\enp}{\hfill\mbox{}\qed}
\makeatletter
\def\moverlay{\mathpalette\mov@rlay}
\def\mov@rlay#1#2{\leavevmode\vtop{%
   \baselineskip\z@skip \lineskiplimit-\maxdimen
   \ialign{\hfil$\m@th#1##$\hfil\cr#2\crcr}}}
\newcommand{\charfusion}[3][\mathord]{
    #1{\ifx#1\mathop\vphantom{#2}\fi
        \mathpalette\mov@rlay{#2\cr#3}
      }
    \ifx#1\mathop\expandafter\displaylimits\fi}
\makeatother

\newcommand{\noi}{\noindent}
\newcommand{\Mi}{\mathbb{M}}
\newcommand{\mghd}{\textsc{mghd}}
\newcommand{\pde}{\textsc{pde}}

\newcommand{\ali}{\begin{align}}
\newcommand{\elin}{\end{align}}

\newcommand{\n}{\nabla}

\newcommand{\p}{\partial}

\newcommand{\GR}{{\sc gr}}

\newtheorem{theorem}{Theorem}
\newtheorem{lemma}[theorem]{Lemma}
\newtheorem{proposition}[theorem]{Proposition}

\newtheorem{definition}[theorem]{Definition}

{ \begin{trivlist}%
  \item[\hskip \labelsep {\bfseries #1}]%
}%
{ \end{trivlist}%
}
\newcommand{\qed}{\nobreak\hfill$\Box$}

\usepackage[symbol]{footmisc}
\renewcommand{\thefootnote}{\fnsymbol{footnote}}
\begin{document}
\pagenumbering{arabic} \setlength{\unitlength}{1cm}\cleardoublepage
\date\nodate
\begin{center}
\begin{LARGE}
{\bf Penrose's 1965 singularity theorem:\\ From geodesic incompleteness to cosmic censorship\footnote[1]{This expository paper is based on my keynote talk \emph{Penrose in perspective} at the conference \emph{Singularity theorems, causality, and all that: A tribute to Roger Penrose (SCRI21)}, June 14--18, 2021, organized by Piotr Chru\'{s}ciel, Greg Galloway, Michael Kunzinger, Ettore Minguzzi, and Roland Steinbauer. I am singularly grateful to these organizers for the invitation to speak. I am also indebted to Michel Janssen for showing me a copy of Einstein's letter to Besso (March 1, 1931) quoted in footnote \ref{bes}, as well as  to Jos\'{e} Senovilla for several important corrections on the first version of this paper. Finally, with  gratitude I spoke to Penrose for two hours on July 2, 2022 in Oxford, followed by a longish walk, which led to further clarifications.}}\end{LARGE}
\bigskip

\begin{Large}
 Klaas Landsman
 \end{Large}
 \medskip
 
 \begin{large}
  Department of Mathematics, Radboud University, Nijmegen, The Netherlands\\
Email:
\texttt{landsman@math.ru.nl}
\end{large}
\smallskip
 \begin{abstract} 
\noindent 
Supplementing earlier literature  by e.g.\ Tipler, Clarke, \& Ellis (1980),  Israel (1987), Thorne, (1994), Earman (1999),  Senovilla \& Garfinkle (2015), Curiel (2019ab), and Landsman (2021ab), I provide a historical and conceptual analysis of Penrose's path-breaking 1965 singularity (or incompleteness) theorem. The emphasis is on the nature and historical origin of the assumptions and definitions used in---or otherwise relevant to---the theorem, as well as on the discrepancy between the (astro)physical goals of the theorem and its actual content:
 even if its assumptions are met, the theorem fails to prove the existence or formation of black holes.
 Penrose himself was well  aware of this gap, which he subsequently tried to overcome with his  visionary and influential cosmic censorship conjectures. Roughly speaking, to infer from (null) geodesic incompleteness that there is 
 a ``black'' object one needs weak cosmic censorship, whereas in addition a ``hole'' exists (as opposed to a boundary of an extendible space-time causing the incompleteness of  geodesics)  if strong cosmic censorship holds. 
\end{abstract}\end{center}
\thispagestyle{empty}
\section{Introduction}\label{HI}
\renewcommand{\thefootnote}{\arabic{footnote}}
 \setcounter{footnote}{0}
 Penrose's short paper `Gravitational collapse and space-time  singularities' from 1965
was surely one of the most important contributions to general relativity since Einstein had formulated the theory in 1915:
\begin{small}
\begin{quote}
However, all bets were off concerning the possibility of a non-singular outcome to a collapse after the publication of a theorem by Roger Penrose (1965), which has claims to be considered the most influential development in general relativity in the 50 years after Einstein founded the theory. Penrose demonstrated that singularities were, after all, a generic feature of gravitational collapse and must appear soon after the formation of a trapped surface (a surface from which light cannot escape outwards). Although the theorem could offer no information about the nature of the singularity, there was a widespread belief that it would have an all-enveloping spacelike character, as in the Schwarzschild case, thus obstructing further development of the situation. Penrose's paper was just as important for what it initiated as for what it accomplished. Powerful global techniques exploiting the causal structure of space-time had been introduced into the theory for the first time. \\ \mbox{} \hfill (Israel, 1987, p.\ 253)
\end{quote}
\end{small}
\begin{small}
\begin{quote}
Penrose's theorem said \emph{roughly} this: Suppose that a star---any kind of star whatsoever---implodes so far that its gravity becomes strong enough to form an \emph{apparent horizon}, that is, strong enough to pull outgoing light rays back inward. After this happens, nothing can prevent the gravity from growing so strong that it creates a singularity. Consequently (since black holes always have apparent horizons), \emph{every black hole must have a singularity inside itself.}\hfill (Thorne, 1994, p.\ 227)
\end{quote}
\end{small}
\begin{small}
\begin{quote}
The singularity theorems constitute the first genuine post-Einsteinian content of classical \GR, not foreseen in any way by Einstein (\ldots) The global mathematical developments needed for the singularity theorems, and the ideas on incompleteness or trapping---and thus also their derived inferences---were not treated nor mentioned, either directly or indirectly, in any of Einstein's writings.
In 1965 \GR\ left adolescence behind, emancipated from its creator, and became a mature physical theory full of vitality and surprises. \hfill (Senovilla \& Garfinkle, 2015, p.\ 2)
\end{quote}
\end{small}
The comparably path-breaking series of papers---this time on the observational side---by the Event Horizon Telescope Collaboration (2019) on their famous image of the black hole in M87 opens as follows:
\begin{small}
\begin{quote}
Black holes are a fundamental prediction of the theory of general relativity (\GR; Einstein 1915). A defining feature of
black holes is their event horizon, a one-way causal boundary in
space-time from which not even light can escape (Schwarzschild
1916). The production of black holes is generic in \GR\ (Penrose 1965). \hfill (Event Horizon Telescope Collaboration, 2019, p.\ L1)
\end{quote}
\end{small}
On the basis of this paper, to general acclaim Penrose won half of the 2020 Physics Nobel Prize:
\begin{small}
\begin{quote}
for the discovery that black hole formation is a robust prediction of the general theory of relativity.
\end{quote}
\end{small}
All records show that this is certainly how the (astro)physics community interpreted 
Penrose's result.\footnote{\label{SrefsFN} Israel (1987), Tipler, Clarke \& Ellis (1980), Earman (1995, 1999),  Senovilla (1998), Senovilla \& Garfinkle (2015), and Curiel (2019a) survey singularities and singularity theorems in \GR\ in a historical context. For textbook treatments see e.g.\ Hawking \& Ellis (1973), O'Neill (1983), Wald (1984),  Clarke (1993),  Beem,  Ehrlich, \& Easley (1996), Senovilla (1998), Kriele (1999), Joshi (2007), and Minguzzi (2019). For Penrose's theorem see also  Aretakis (2013) and Galloway (2014).
\label{refsfn}} In particular, it put an end to claims even by leading relativists to the effect that curvature singularities of the kind found in the static Schwarzschild solution from 1916 (and later in the Kerr solution from 1963), as well as in the dynamical Oppenheimer--Snyder solution from 1939, were artificial consequences of unphysical idealizations (such as spherical or axial symmetry) and would be absent in realistic models. 

 But let us recall  that Penrose (1965c) \emph{actually} proved that in the presence of a suitable curvature condition,
  a space-time  $(M,g)$ that  contains a non-compact Cauchy surface as well as a  trapped surface cannot be
  future null geodesically complete. Thus it is far from obvious how Penrose's  theorem relates to the above summaries by Thorne, the \textsc{eht}, and the Nobel Prize committee (Israel is more accurate; the discontinuity between Einstein and Penrose claimed by Senovilla and Garfinkle will be addressed).

Putting it bluntly, the main differences between Penrose's actual theorem and its reputation are that:
\begin{itemize}
\item  The theorem says \emph{nothing} about event horizons, which form the ``black'' ingredient of a black hole;
\item It is also \emph{inconclusive} about ``singularities'', which should form the ``hole'' part of a black hole.\end{itemize}
Hence the link between the theorem and black holes  is opaque. What is missing is the further inference, perhaps under additional assumptions, from the existence of incomplete null geodesics to the existence of an event horizon, as well as an argument relating  incompleteness to the existence of a singularity in some physically meaningful sense (such as a curvature singularity), as opposed to, for example, the sudden end of a geodesic being caused by  extendibility of the space-time  under consideration. Penrose (1965c) was well aware of both problems, see \S\ref{exegesis} below. Addressing the first  had to wait for a good definition of an event horizon, which he provided in Penrose (1969) and which led, in the same paper, to what is now called his \emph{weak cosmic censorship} conjecture. Penrose subsequently addressed the second (extendibility) problem, albeit rather indirectly at first, by what we now call  \emph{strong cosmic censorship}.

 Here is the plan of the remainder of this paper:
 \begin{description}
 \item[]
 \S\ref{before}  summarizes the status of singularities and singularity theorems in \GR\ before Penrose (1965c).\footnote{This section adds only little to Tipler, Clarke, \& Ellis (1980), Earman (1999), and   Senovilla \& Garfinkle (2015). }
 \item[]  \S\ref{exegesis} goes through Penrose (1965c) in detail, paying special attention to what was and was not new.
 \item[] \S\ref{intermezzo} resumes the discussion of singularities, including Penrose's (1974, 1979) ``localization'' thereof.
  \item[]\S\ref{cosmic} discusses cosmic censorship mainly from the point of view of
  Penrose (1969, 1973, 1974, 1979).
     \item[]\S\ref{future} explains the transition from 
  his cosmic censorship conjectures to the versions that are currently used.
  \item[] \S\ref{epilogue} is an Epilogue, reflecting a little on  scientific revolutions and historiography in the present context.
 \end{description}
Let me stress, perhaps against first appearances, that this paper is intended to \emph{increase} our admiration of Penrose's work in \GR, which---apart from equally deep and influential ideas on gravitational radiation and asymptotic flatness, spinor and twistor theory,  black hole thermodynamics, and cosmology---may be seen as a sustained effort to accomplish what many people seem to believe  his 1965 theorem already did. 
 \section{Singularities and singularity theorems before 1965}\label{before}
 The period from 1916 to 1965 is well covered by Tipler, Clarke, \& Ellis (1980), Earman (1995, 1999),  Earman \& Eisenstaedt (1999), and Senovilla \& Garfinkle (2015), but problems of interpretation remain. Like the history of \GR\ itself (Janssen, 2014; Janssen \& Renn, 2022), I see a steady but bumpy road---this time from Einstein to Penrose---where ultimately the mathematician overruled the physicist and won.
 
Both Hilbert (1917) and Einstein (1918) saw singularities as points \emph{in space-time } where some components $g_{\mu\nu}$ of
   the metric were ill-defined and/or $\det(g)=0$, with some  attention paid to the role of the coordinate system. Such singularities were seen in the Schwarzschild solution from 1916 at  $r=2m$, as well as in the de Sitter solution from 1917.\footnote{The Friedman solution  (rediscovered by Lema\^{i}tre in 1927) dates from 1922, so it was not part of the earliest discussions.}  Slightly rewritten by Einstein (1918), de Sitter's metric is
 \begin{equation}
ds^2=dr^2+R^2\sin^2\left(\frac{r}{R}\right)d\Om^2
-\cos^2\left(\frac{r}{R}\right) dt^2,\label{dS}
\end{equation}
where $d\Om^2=d\theta^2+\sin^2\theta d\phv^2$, and
$R$ is related to the cosmological constant $\lm=3/R^2$ (de Sitter, 1917). 
Einstein's understanding of singularities has (understandably) been questioned,\footnote{For example: `He tended to be more
disturbed by (what today we would call) merely apparent singularities and less
disturbed by (what we would call) real singularities. Einstein had strong a priori
ideas about what results a correct physical theory should deliver. (\ldots) He tended to push aside technical
problems and jump over essential diffculties. (\ldots) His difficulties lay with cases that today beginning students of \textsc{gtr} take for granted.' (Earman \& Eisenstaedt, 1999, p.\ 185)}
 but let me note that:
\begin{itemize}
\item  Not just Einstein but even Hilbert (1917), who was also an expert in \GR, failed to recognize the apparent ``singularity'' of the Schwarzschild metric at $r=2m$ as a mere coordinate singularity.\footnote{And they were not the only ones:  `until the early 1960s the general opinion was that it was a real singularity' (Earman \& Eisenstaedt, 1999, p.\ 188).
 Hilbert (1917) required a coordinate transformation intended to remove an apparent singularity 
to be regular (i.e.\ continuous and continuously differentiable)  and invertible even at the location of the (alleged) singularity. This defeats the purpose and is not even true for changing from Cartesian  to polar coordinates. `How Hilbert, one of the great mathematical minds of the century, could have failed to appreciate  this elementary point defies rational explanation.' (Earman \& Eisenstaedt 1999, p.\ 193).
The sphere at the Schwarzschild radius $r=2m$ was alternatively described as a ``discontinuity'' (Schwarzschild), ``magic circle'' (Eddington), ``barrier'' (Kottler), ``limit circle'' (Brillouin), and even ``the death'' (Nordmann); I learnt this in a seminar by Dennis Lehmkuhl on April 12th, 2021.  After important but inconclusive work by
  Eddington (1924),
 Lema\^{i}tre (1933), \S 11, correctly concluded that `The singularity of the Schwarzschild field [i.e.\ at $r=2m$] is thus a fictitious singularity, analogous to that which appears at the horizon of the centre in the original form of the de Sitter universe.' However, his work was not well known and took decades to be digested. For example, Einstein (1939), p.\ 922, apparently unaware of  Lema\^{i}tre (1933), still  called the Schwarzschild metric ``singular'' at $r=2m$ since $g_{00}$ vanishes and `both light rays and material particles take an \emph{infinitely} long time \emph{(measured in ``coordinate time'')} in order
to reach the point $r=2m$ when originating from a point $r>2m$' (scare quotes are Einstein's but emphasis added to highlight the changes from his reply to de Sitter). See also Godart (1992), Eisenstaedt (1993),  Earman (1995), \S 1.2,
Earman \& Eisenstaedt (1999), \S 2,  and van Dongen \& Lehmkuhl (in progress).
 It seems that
  Finkelstein (1958) was the first author to introduce the modern understanding of $r=2m$ (now no longer seen as singular) as a one-way membrane.  But he did not use the term ``event horizon'', which had just been introduced by Rindler (1956), who in turn did not mention the Schwarzschild solution!  It was Penrose (1968) who first  put all this together. 

Until the 1930s almost no attention was paid to what we now understand to be the real singularity of the Schwarzschild metric at $r=0$, firstly because Newtonian gravity has a similar singularity seen as unproblematic, and secondly because  
this vacuum solution was not supposed to be valid in the interior of a star, i.e.\ as $r\raw 0$; indeed no star was known or expected to have a radius $r<2m$. This was also Einstein's (final) reply to Hadamard, who
during a lecture in Paris on April 5, 1922,  asked Einstein what  happened if the radius of the Sun were less than the Schwarzschild radius (Biezunski, 1987).}
\item
 Not just Einstein but even Weyl (1918)  saw  $r=\pi R/2$ as a genuine singularity in de Sitter's metric.\footnote{See Janssen (2014), pp.\ 202--207, for this episode. de Sitter space-time, independently introduced by
 Levi-Civita (1917), has no singularities at all (it is also globally hyperbolic), although it is not geodesically convex (Calabi \& Markus, 1962). 
 } But Einstein was not confused by the coordinate singularity at $r=0$ and did understand
the  need for different coordinate patches in order to define regularity (and hence singularity) of the metric:
\begin{small}
\begin{quote}
Furthermore, the continuity condition for the $g_{\mu\nu}$ and the $g^{\mu\nu}$ [including their first derivatives, as he says earlier] must not be interpreted to mean that it must be possible to choose coordinates such that the conditions are satisfied in the entire space. Obviously, one must only demand that for the neighborhood of every point one can select coordinates such that the continuity conditions are met within this neighborhood. \mbox{  } (Einstein, 1918, p.\ 271; Einstein, 2002, p.\ 37).
\end{quote}
\end{small}\end{itemize}
Counterfactually granting Einstein (1918) that  $r=\pi R/2$ is indeed a ``singularity''---since he believed that `no choice of coordinates can remove this discontinuity'---the interesting point is his subsequent argument that it is `genuine'.
Namely, he defines a point $P$ to lie `in the finite domain when it can be connected by a curve with a fixed, chosen point $P_0$, so that the distance integral $\int_{P_0}^Pds$ has a finite value.' He then shows that the (alleged) singularities at $r=\pi R/2$ lie `in the finite domain', concluding that `the De Sitter solution has a genuine singularity'
 (\emph{eine echte Singularit\"{a}t}'). This, then, he
 considers  `a grave argument against the admissibility of [de Sitter's] solution'. Thus Einstein's logic seems crystal clear:
 \begin{itemize}
\item  He first defines precisely what he means by a singular space-time  (or `solution', as he calls it):
\begin{enumerate}
\item There must be at least one point $P$ where in all coordinate systems  some component $g_{\mu\nu}$ or $g^{\mu\nu}$ (or its first derivative) fails to be continuous (this is the case, for example, if $\det(g)=0$);
\item Such points $P$ can be connected to ``regular'' points $P_0$ by some curve of finite proper length.
\end{enumerate}
\item Admissible space-times must be non-singular, i.e., either the metric is regular at all points $P$, or, if there are any singular points, these must lie at infinite (proper) distance from all regular points.
\end{itemize}
Criterion 1 is at odds with the modern notion of a (smooth) Lorentzian manifold, which cannot contain such points. But this was far from obvious at the time and still needed to be spelled out as late as 1963:\footnote{Even for mathematicians, leaving out singular points from the space under consideration is not  an obvious move, as for example algebraic geometry shows. Within differential geometry, stratified spaces provide another example (Pflaum, 2001).}
\begin{small}
\begin{quote}
the clue to clarity is to refuse ever
to speak of a singularity but instead to phrase
everything in terms of the properties of differentiable
metric fields on manifolds. If one is given a manifold,
and on it a metric which does not at all points
satisfy the necessary differentiability requirements,
one simply throws away all the points of singularity.
The starting point for any further discussion is
then the largest submanifold on which the metric is
differentiable. This is done because there is not
known any useful way of describing the singularities
of a function except by describing its behavior at
regular points near the singularity. The first problem
then is to select a criteria which will identify in an
intuitively acceptable way a ``nonsingular space.''
Evidently, differentiability is only a minimum
prerequisite, since everything becomes differentiable
when the singular points are discarded. The problem
is rather to recognize the holes left in the space
where singular (or even regular) points have been
omitted. \hfill (Misner, 1963, p.\ 924)
\end{quote}
\end{small}
I will discuss Misner's paper below. 
Returning to Einstein (1918), at least in a Whig interpretation of history his second criterion may be claimed to foreshadow geodesic incompleteness. It would have been more reasonable, though,  if he had used \emph{causal geodesics} instead of \emph{arbitrary curves}; first deciding for \emph{causal} curves on  physical grounds and/or because this is enough for all interesting examples,\footnote{In de Sitter space-time  one can reach singular points  ($r=\pi R/2$) from regular ones  ($0\leq r<\pi R/2$) with any kind of curve.} his criterion does not even make sense if arbitrary (as opposed to geodesic) causal curves are allowed.\footnote{First, any two points that can be connected by a lightlike (i.e.\ null) curve can also be connected by either a 
lightlike \emph{geodesic} or a timelike \emph{curve}; see e.g.\ Minguzzi (2019), Theorems 2.20 and 2.22, or Landsman (2021a), Proposition 5.13, all going back to Penrose (1972). Second, as pointed out by Clarke (2003, pp.\ 2--3) in precisely the present context of Einstein (1918),  any two points that can be connected by a timelike curve can also be connected by a timelike curve \emph{of finite proper length} (namely by `wiggling the curve to make its speed close to the speed of light'), which makes Einstein's criterion vacuous as stated. See also Earman \& Eisenstaedt (1999) and Nussbaumer \& Bieri (2009) for broader discussions of this episode.
}

Even in the absence of satisfactory definitions of a singular space-time, towards which little progress was made until the 1960s, one may look for ``singularity theorems'' prior to Penrose's. But  because it is not clear what is meant by a ``singularity'', there is considerable leeway in identifying such theorems, ranging from accepting Genesis 1:1 to rejecting anything before Penrose (1965c)  to deserve the name.\footnote{Following the Friedman solution from 1922 and its rediscovery by Lema\^{i}tre (1927), who---unlike the mathematician Friedman---as an astronomer
  related the solution to the redshifts of extragalactic nebulae that had been discovered in 1917,
it was well understood in the 1930s that theoretically speaking the expanding universe might have had a beginning, but this possibility was not taken very seriously until the 1960s (Tipler, Clarke, \& Ellis, 1980). See  Nussbaumer \& Bieri (2009),  Nussbaumer (2014), and Smeenk (2014) for Einstein's views. Believing the universe was static, his initial reaction to Friedman (1922) and Lema\^{i}tre (1927) was that their solutions were `irrelevant' and `abominable', respectively, but after talking to Eddington in 1930 and Tolman in 1931 he came to accept an expanding universe. As a letter to Besso from March 1, 1931 shows, Einstein also clearly realized, like many others, that `the expansion of matter points to a beginning of time which lies $10^{10}$, respectively $10^{11}$ years in the past.' (`Der Hacken ist aber, dass die Expansion der Materie auf einen zeitlichen Anfang schliessen l\"{a}sst, der $10^{10}$ bzw.\  $10^{11}$ Jahre zur\"{u}ckliegt').
The so-called Einstein--de Sitter universe (which is a flat \textsc{flrw} model remaining popular until the end of the twentieth century) indeed starts from what we now call a big bang, but
 like other leading relativists including de Sitter (!) and Eddington,  Einstein tried to avoid this conclusion, in his case by   
`pointing out that the inhomogeneity of the distribution of stellar material makes our approximate treatment illusory'; in other words, by appealing to the unphysical idealizations inherent in the symmetries of the Friedman--Lema\^{i}tre(--Robertson--Walker) solution. As we will see, this became a common move against singularities in \GR, which Penrose (1965c) and  Hawking (1965, 1966ab)  made obsolete. \label{bes}
 } 
 
 \noi
 For example,  both Tipler, Clarke, \& Ellis (1980)
and  Earman (1999) attribute the first singularity theorem to Tolman \& Ward (1932),  whereas Senovilla \& Garfinkle emphatically start with Raychaudhuri (1955). 

Prior to Robertson (1933) and Walker (1937), Tolman \& Ward (1932)  analyzed closed homogenous and isotropic universes filled with matter in the form of a perfect fluid with $\rho>0$ and $p\geq 0$ (and zero cosmological constant), to conclude that such models have a beginning in a state of zero volume and also recollapse after a state of maximal expansion. However, they did not conclude  that there was a big bang, instead believing in some sort of cyclic process of repeated expansions, recollapses, and bounces.\footnote{From a modern point of view such a process means that their space-time can be extended through $t=0$, which in the case at hand is impossible because of the curvature singularity as $t\raw 0$ (in this case it is the Ricci scalar that diverges).}

Earman (1999) reviews various analogous cosmological results from the 1930s, due  to 
Tolman, Robertson, de Sitter, Synge, and Lema\^{i}tre,
 which with hindsight may be classified as ``singularity theorems'', in which ``singularities'' are defined in diverse ways, including zero volume (or radius), infinite density, as well as problematic expressions for the metric.
 Lema\^{i}tre (1933), which stands out in its astronomy, mathematics, and physics,  was the first paper to use an energy condition.\footnote{This happens after eq.\ (2,10) on page 678 of the reprint in 1997, where he states that `in all reasonable applications $T_1^1$, $T_2^2$ and $T_3^3$ will be negative, and in all cases less than $T^4_4=\rho$ in absolute value'.} But all these authors immediately give arguments why their results \emph{must} be mathematical artifacts, typically by appealing to their unrealistic  underlying idealizations or to unknown new physics. Even  Lema\^{i}tre (1933)   
writes: 
 \begin{small}
\begin{quote}
The matter has to find, though, a way of avoiding the vanishing of its volume.  (p.\  678 of 1997 reprint)
 \end{quote}
\end{small}

Meanwhile, progress was also made on the astrophysical side.\footnote{See Israel (1987), Thorne (1994), and Longair (2006) for the relevant history, which is fascinating by itself.} Light stars were found to retire as white dwarfs, in which nuclear burning has ended and inward gravitational pressure is stopped by a degenerate electron gas. In 1931, Chandrasekhar and Landau  independently discovered that this only works for masses $M\leq 1.46 M_{\odot}$ (I take the current value); heavier stars collapse into neutron stars (typically after a supernova explosion). But also these have an upper mass, as first suggested by Oppenheimer \& Volkoff (1939); the current upper bound is about $2.3  M_{\odot}$. Heavier stars collapse into a black hole.

Yet this conclusion, so obvious with hindsight, was resisted as much as the idea of a big bang, despite a paper by Oppenheimer \& Snyder (1939) in which the collapse process was explicitly described within \GR, albeit under the assumption of spherical symmetry and an idealized equation of state.\footnote{As pointed out to me by Senovilla, the  interior part of their solution is  the contracting part of Friedmann's (1922) solution. See also Datt (1938), whose work was all but  ignored; the poor author `died soon after on the operation table' (Dadhich, 2020).}
 It was:
\begin{small}
\begin{quote}
a manuscript that has strong claims to be considered the most daring and uncannily prophetic paper in the field. There is nothing  in this paper which needs revision today. \hfill (Israel 1987, pp.\ 226--227)
\end{quote}
\end{small}
But there is a big difference between our perspective and the contemporary one (see also \S\ref{epilogue}):
\begin{small}
\begin{quote}
At the end of the 1930s not only was there no agreement on how to define singularities, there was not even a consensus about the status of singularities in the key test case, the Schwarzschild solution. \\ \mbox{}  \hfill
(Earman, 1999, p.\ 240)
\end{quote}
\end{small}
Restricting this summary to events directly relevant to Penrose's paper, all authors agree that the next key paper was Rauchaudhuri (1955); as already mentioned, Senovilla \& Garfinkle (2015) even take this to contain the very first singularity theorem in \GR. Rauchaudhuri studies a universe filled with dust, i.e.\footnote{Raychaudhuri's  reasoning  breaks down if a pressure term $ph_{\mu\nu}$ is added to $T_{\mu\nu}$, since in that case the congruence he uses is no longer geodesic, which in turn adversely affects his analogue of the central equation \er{RE}. }
\beq
T^{\mu\nu}=\rho u^{\mu}u^{\nu}, \label{dust}
\eeq
 but allows $\rho$ to be an arbitrary positive function $\rho(t,\vec{x})$, as opposed to previous treatments where $\rh$ depends on $t$ alone. Indeed, the goal of his paper is clear from its abstract, which states that:
 \begin{small}
\begin{quote}
a simple change over to anisotropy without the introduction of spin does not solve any of the outstanding difficulties of isotropic cosmological models.\footnote{By `spin' Raychaudhuri means $\omega_{ij}=\half(\n_iu_j-\n_ju_i)$, noticing that $\om_{ij}=0$ iff the flow $u^{\mu}$ is hypersurface orthogonal. He cites G\"{o}del (1949) as an example where `spin' contributes to making this famously acausal solution nonsingular. G\"{o}del's paper also played a significant role in the mathematical turn \GR\ took in the 1960s (Tipler, Clarke, \& Ellis, 1980, p.\ 111).
The July 2007 issue of \emph{Pramana--Journal of Physics} (Volume 69, no.\ 1) is devoted to Raychaudhuri and his equation. See also Tipler, Clarke, \& Ellis (1980), 
Earman (1999), and Ehlers (2006) for the history and context of Raychaudhuri's work in \GR. }  
  \hfill (Rauchaudhuri, 1955, p.\ 1123)
\end{quote}
\end{small}
Instead of the parameter $\theta=\n_{\mu}u^{\mu}$ familiar from modern renditions of his work, Raychaudhuri uses the volume element $\sqrt{-\det(g)}$, or rather $G=(-\det(g))^{1/6}$, but these are related.\footnote{If $u^{\mu}$ is hypersurface orthogonal and normalized to $g_{\mu\nu}u^{\mu}u^{\nu}=-1$, the relation is $\theta =\p_t\ln(\sqrt{-\det(h)})$, where $h$ is the metric induced by $g$ on the  hypersurface to which the flow is orthogonal.
See e.g.\ Landsman (2021a), \S 6.1, eq.\ (6.25).} 
 For this reason, the famous equation for $\theta$ named after him, namely (assuming that $\n_uu=0$, which follows from \er{dust})
 \beq
 \n_u\theta\equiv\dot{\theta}=-\third\theta^2-\sg_{\mu\nu}\sg^{\mu\nu}+\om_{\mu\nu}\om^{\mu\nu}-R_{\mu\nu}u^{\mu}u^{\nu}, \label{RE}
 \eeq
cannot be found in his paper;\footnote{Here $\sg_{\mu\nu}:=k_{(\mu\nu)}-\third\theta h_{\mu\nu}$, with $k_{\mu\nu}=h^{\rh}_{\mu}h^{\sg}_{\nu}\n_{\rh}u_{\sg}$, where $h$ is the spatial metric $ h_{\mu\nu}=g_{\mu\nu}+u_{\mu}u_{\nu}$, and
$\om_{\mu\nu}=k_{[\mu\nu]}$. Eq.\ \er{RE} appears, for example, in Ehlers (1961), eq.\ (36), which is the earliest source I know.
} instead, one finds an equivalent (but less transparent) second-order differential equation for $G$. Either way, if $R_{\mu\nu}u^{\mu}u^{\nu}\geq 0$ (which Raychaudhuri infers from the energy condition $\rho\geq 0$ and the Einstein equations) and $\om_{\mu\nu}=0$ (i.e.\ if the flow is irrotational), then,
since $\sg_{\mu\nu}\sg^{\mu\nu}\geq 0$ for any congruence,
 \er{RE} implies that if $\theta>0$ on some hypersurface orthogonal to the flow at some time, i.e.\ the universe is expanding, then $\theta\raw\infty$ at some finite time in the past, which corresponds to $G\raw 0$. 
Since $\rh\sqrt{-\det(g)}=\rh G^3$ is constant, this suggests  $\rho\raw \infty$, but $G$ depends on the coordinates and  $G\raw 0$ could merely mark a singularity of the congruence.
 \emph{On
 Raychaudhuri's assumption that matter moves along the timelike geodesics tangent to $u$}, however, he does obtain a   density and  curvature singularity, adding
 that the time from such a `singular state' to `the present state' is a \emph{maximum} for isotropic models, so that adding anisotropy to  isotropic cosmological models even brings the initial singularity forward!\footnote{In general, $\theta\raw\infty$ could indicate that the hypersurface orthogonal to the flow  becomes lightlike (and then timelike), in which case neither the density nor the space-time needs to be singular (Ryan \& Shepley, 1975; Collins \& Ellis, 1979; Tipler, Clarke, \& Ellis, 1980; Earman, 1999). 
 Instead of a singularity in the energy density this  implies that the domain of influence of some initial spacelike hypersurface carrying the initial data for the Einstein equations comes to an end, so that a (necessarily lightlike) Cauchy horizon forms.  This  happens, for example, in the Taub--NUT space-time (Misner, 1963; Shepley, 1964; Misner \& Taub, 1969). Cauchy horizons may still lead to singularities under a perturbation that destabilizes the Cauchy horizon, cf.\ \S\ref{future}. 
This cannot happen in Bianchi type I models (Heckmann \& Sch\"{u}cking, 1962) and type IX models (Shepley, 1964), where  Raychaudhuri's argument does lead to a singularity in the energy density and even in space-time. }

 Similar comments apply to Komar (1956), whose Introduction is worth quoting as a sign of the times:
  \begin{small}
\begin{quote}
A principal success of the general theory of
relativity in the realm of cosmology is given by
the Friedmann solution of the field equations. This solution, which employs the assumptions that the universe is spacially isotropic and that the state of matter may be represented by incoherent dust, yields the result that the universe is not stationary, but is rather in a state either of expansion from a singular point in time (which would correspond to creation), or of contraction toward a singular point in time (which would correspond to annihilation). The question naturally arises whether such singular points are a consequence of the particular symmetry presupposed in Friedmann's model, or whether perhaps for more general distributions of matter one need not expect instants of creation or annihilation of the universe. The purpose of this paper
is to show that singularities in the solution of the field equations of general relativity are to be expected under very general hypotheses (enumerated specifically below), and in particular that the singular instant of creation (or annihilation) necessarily would occur at a finite time in the past (or future, respectively). \hfill (Komar, 1956, p.\ 104)
\end{quote}
\end{small}
Replacing `cosmology' by `gravitational collapse' and `Friedmann' by `Schwarzschild', this purpose is hardly different from Penrose's: it echoes the general spirit  that one would like to go beyond the simplified singular models of the past and see if their singularities persist (which Komar shows they do).

 In modern fashion, Komar just relies on energy assumptions, due to which he need not specify $T_{\mu\nu}$ explicitly---unlike 
 Raychaudhuri (1955), whom he does not cite.\footnote{Raychaudhuri (1957) seems a little reprimand for this. Senovilla (1998) combines their results in his Theorem 5.1.} These are:
 $T_{00}\geq 0$ and $T\geq 0$, and $T_{00}=0$ implies $T_{ij}=0$. But by \emph{fiat} he forces his geodesic congruence to be the same as Raychaudhuri's, including its identification with the matter flow lines. This then leads him to very similar conclusions.\footnote{Similar results were found by Landau, whose Russian colleagues recognized their inconclusiveness (Earman 1999, \S 5).
This eventually led to the so-called BKL-approach to singularities in \GR. See Belinski \& Henneaux (2018) for a recent review.
}  

Both papers crossed mathematical work on relativistic fluid mechanics,\footnote{For example, prior to 1955 by  Eisenhart, Synge, Lichnerowicz, and subsequently by the Hamburg group in \GR\ founded by Jordan, consisting notably of Behr, Ehlers, Heckmann,  and Sch\"{u}cking. See Tipler, Clarke, \& Ellis (1980) for references.
 } which in turn crossed or included the study of 
 spatially homogeneous (Bianchi) cosmologies initiated by Taub (1951), G\"{o}del (1952), and the Hamburg group, summarized by Heckmann \& Sch\"{u}cking (1962). The general goal  was 
   \begin{small}
\begin{quote}
 to find singularity free [!] fluid filled Bianchi cosmological models. \hfill (Ellis, 2014, p.\ 2)
 \end{quote}
\end{small}
  This research influenced all subsequent developments, where  Raychaudhuri's  argument survived as a lemma for the mature singularity theorems, in whose proofs it is used to
 prove focusing of geodesics.
 
 To preface the next phase involving the young relativists Misner, Penrose, and Hawking, I quote:
  \begin{small}
\begin{quote}
Wheeler and Sciama had a widespread impact on the development of the understanding of singularities, particularly through their students. Wheeler's students included Misner, Shepley, Thorne, and Geroch, while Sciama's included Ellis, Hawking, Carter, Rees, and Clarke. However, Sciama has claimed, on occasion, that his most important contribution to relativity has been in influencing Roger Penrose to work on the subject! \hfill (Tipler, Clarke, \& Ellis, 1980, p.\ 136)
\end{quote}
\end{small}
The last pre-Penrose papers on singularities I mention are Misner (1963) and  Shepley (1964, 1965).\footnote{Misner (1963), footnote 4, thanks his PhD student Shepley for `preparing this review' (i.e.\ of the definition of a singularity in space-time in the Introduction), adding that `We have borrowed heavily from [lectures by]  L. Marcus' (sic),  i.e.\ Lawrence Mar\ul{k}us (1922--2020). Shepley (1965) is a Princeton PhD thesis  supervised by Misner and Wheeler. Some relevant passages reappear in Ryan \& Shepley (1975),  from which I will quote as if these were written in 1963--1964.
Footnote 19 in Misner (1963) shows that Misner \& Taub (1969), which contains a similar analysis, must have been almost ready already in 1963.\label{Misnerfn}
} 
Once again, but for a different reason, it is instructive to quote from the Introduction of Shepley (1964):
  \begin{small}
\begin{quote}
Most cosmological models in general relativity have a point of singularity (\ldots) which can be reached by a geodesic of finite total length from other
points of the space-time manifold, where the metric is degenerate or otherwise
irregular (for example, a point where a curvature scalar is infinite). All known
dust-filled models, without cosmological constant, have singular points, and it
may be conjectured that the presence of dustlike matter filling space always leads
to a singularity. This paper shows that this conjecture is true for an important
class of universes.   \hfill (Shepley, 1964, p.\ 1403)
\end{quote}
\end{small}
This  shows that using geodesic incompleteness as a characterization of a singularity was familiar at the time, although e.g.\ Heckmann \& Sch\"{u}cking (1962) do not mention it.\footnote{Mentioning \GR\ but not  the search for a definition of a  singular space-time, Kundt (1963), Hermann (1964), and Fierz \& Jost (1965) all study causal geodesic (in)completeness in Lorentzian manifolds in comparison with the Riemannian case.}
 But it was seen as insufficient:
  \begin{small}
\begin{quote}
completeness is vulgarly used as the definition of non-singularity. \hfill (Ryan \& Shepley, 1975, p.\ 80)
\end{quote}
\end{small}
Thus Misner (1963)  refines this idea.
After the call to arms already quoted in \S\ref{before}, followed by a comparison with the Riemannian case,    geodesic incompleteness is proposed as a \emph{necessary} condition for  a space-time to be singular, in which case it is \emph{sufficient}  that `some scalar polynomial in
the curvature tensor and its covariant derivatives be unbounded on an (open) geodesic segment of
finite length,' since this implies inextendibility of the geodesic in question into some potential extension of the space-time.\footnote{ For a detailed treatment of all of this see Beem,  Ehrlich, \& Easley (1996), Chapter 6. See also Chru\'{s}ciel,  (2020), \S 4.4. If each finite segment of some incomplete geodesic is contained in a compact subset of space-time, then Misner does not seem to count  this incompleteness as a singularity, since in that case space-time is typically extendible (as happens in  Taub--NUT).} 

Shepley (1964) claims  that Bianchi type IX universes filled with dust are  singular even if the congruence of matter flow lines is rotating.  But although he finds points where  $\det(g)=0$, he
 shows neither geodesic incompleteness, nor divergence of the density or of some
curvature scalar, admitting that
`The question of what happens as the singular point approaches is presently under investigation' (p.\ 1409).\footnote{This may refer to Matzner,   Shepley, \& Warren (1970), which is criticized in Collins \& Ellis (1979), 
  or to Ryan \& Shepley (1975), Chapter 10, which at last proves geodesic incompleteness and inextendibility. See also Earman (1999).}
 \section{Gravitational collapse and space-time singularities: Exegesis  }\label{exegesis}
 The brevity   (2.5 pages) and brilliance of Penrose (1965c) makes almost every comma relevant, but let me try some restraint. Following a brief discussion of gravitational collapse, Penrose explains his motivation:
   \begin{small}
\begin{quote}
The question has been raised as to whether
[the Schwarzschild] singularity is, in fact, simply a property of the high symmetry assumed. The matter collapses radially inwards to the single point at the center, so that a resulting space-time catastrophe there is perhaps not surprising. Could not the presence of perturbations which destroy the spherical symmetry alter
the situation drastically? (\ldots)

 It will be shown that, after a certain critical condition has been fulfilled, deviations from spherical symmetry cannot prevent space-time singularities from arising. \hfill (Penrose, 1965c, p.\ 57, 58)
\end{quote}
\end{small}
 Penrose does not strictly define what he means by a singularity, and cites none of the papers discussed in the previous section except Oppenheimer \& Snyder (1939). But the examples he gives (namely gravitational collapse, Schwarzschild, and Kerr) suggest that he has curvature singularities in mind. Indeed, the preamble to his theorem shows that he does \emph{not} identify singularities with incomplete geodesics:\footnote{I suppose  `completeness' here means \emph{inextendibility} as opposed to \emph{geodesic completeness}, since a few lines later Penrose states his theorem  as the incompatibility between certain assumptions (see main text) and 
future null completeness, which is subsequently specified as `Every null geodesic in $M_+^4$ can be extended into the future to arbitrarily affine parameter values (null completeness)' i.e.\ null \emph{geodesic} completeness.
 Similarly, (c) above only makes sense if Penrose takes `incomplete' to mean \emph{extendible}.
 I thank Erik Curiel and Jos\'{e} Senovilla for a discussion of this point (via email, May 2021), about which we agree.
} 
   \begin{small}
\begin{quote}
The existence of a singularity can never be inferred, however, without an assumption such as completeness [i.e.\ inextendibility] for the manifold under consideration. \hspace{95pt} (p.\ 58)
\end{quote}
\end{small}
Similarly, in a list of four possibilities to avoid singularities despite his theorem,\footnote{The others are:
`(a) Negative local energy occurs; (b) Einstein's equations are violated; (d) The concept of space-time loses its meaning at very high curvatures---possibly because of quantum phenomena.' Penrose omits (e): There is no Cauchy surface. But,  as Senovilla pointed out to me, on a modern understanding this may be taken to be a consequence of his (c).}
 Penrose includes:
   \begin{small}
\begin{quote}
(c) The space-time manifold is incomplete [i.e.\ extendible].\footnote{ A footnote at this point adds that `The ``I'm all right, Jack'' philosophy with regard to the singularities would be included under this heading!'. This reconfirms that  Penrose considers extendibility a possible but cheap way to avoid singularities.} \hfill (p.\ 58)
\end{quote}
\end{small}
\emph{Geodesic} completeness first appears in the statement of the theorem, in which it is a \emph{reductio ad absurdum} assumption, see below, whose negation as a way out of the ensuing contradiction is left to the reader and is nowhere defined as a singularity.  The logic of the above two quotes therefore seems to be that inextendibility of the manifold, as opposed to a `physical' singularity, could be the cause of geodesic incompleteness and as such has to be excluded to make room for a curvature singularity as the physically relevant cause of geodesic incompleteness. This logic foreshadows strong cosmic censorship; cf.\ \S\ref{cosmic}.
 
We now come to the theorem, which states that the following assumptions are together inconsistent:
\begin{theorem} \label{P65theorem} For any  space-time  $(M,g)$, the following assumptions are contradictory:\footnote{In the preamble to this list Penrose supposes that $M$  is `the future time development' of $C_3$. Short of anything like the results of Choquet-Bruhat \& Geroch (1969), this simply seems to refer to his definition of $C_3$ as a Cauchy surface, see below.}
 \begin{enumerate}
 \item A space-time is a four-dimensional time-orientable Lorentzian manifold;\footnote{Penrose always works in signature $(+---)$ and includes a cosmological constant $\lm$ in condition 4, which I put to zero.}
\item $(M,g)$ is future null geodesically complete;
\item $M$ contains a non-compact Cauchy surface $C_3$;
\item One has $(-R_{\mu\nu}+\half g_{\mu\nu} R)t^{\mu}t^{\nu}\geq 0$ for any timelike vector $t$ (at any point);\footnote{For null vectors $t$ this inequality implies $R_{\mu\nu}t^{\mu}t^{\nu}\leq 0$
by continuity. This is what is actually used in the proof. Penrose (1968) indeed states the curvature condition in the latter way. As we have seen, such conditions were hardly new in 1965, neither as curvature conditions nor, via the Einstein equations, as energy conditions (which here would simply be $T_{\mu\nu}t^{\mu}t^{\nu}\geq 0$). }
\item There exists a trapped surface in $M$.
\end{enumerate}\end{theorem}
Penrose (1965c) defines a Cauchy surface by
the property that every inextendible timelike or null geodesic meets it.\footnote{On the tacit understanding that $C_3$ is achronal (or semispacelike, in Penrose's own terminology) this is equivalent to the property $D(C_3)=M$ (where the domain of dependence is defined via \emph{timelike} curves), and hence to the  later definition to the effect that $C_3$ is  any subset of $M$ that intersects every  inextendible timelike curve  once. See Penrose (1968, 1972).  Penrose (1965a) had already loosely defined a Cauchy surface $C_3$ as a hypersurface that intersects every null geodesic exactly once. } This specific concept of a Cauchy surface in \GR\ seems to have originated with Penrose, who therefore, with Leray (1953), Choquet-Bruhat (1968) and Geroch (1970a) should be included on the list of  architects of global hyperbolicity in \GR.\footnote{Domains of dependence $D(S)$ of hypersurfaces $S\subset M$, and the closely related concept of a Cauchy surface (for which $D(S)=M$)
 originate in hyperbolic \pde\ theory in flat space, for which Courant \& Hilbert 1962) was the standard reference at the time.
The equivalence between the existence of a Cauchy surface as Penrose (1965c) defines it and global hyperbolicity as used in the analysis of the Cauchy problem for \GR\ by Leray (1953) and Choquet-Bruhat (1968), i.e.\ compactness (in a suitable topology) of the space of causal curves $C(x,y)$ for any pair  $(x,y)$ of causally connected points, is announced in Penrose (1968, p.\ 191), based on lectures in 1967 (a footnote on the same page says that `This will be shown elsewhere'). Penrose even mentions the factorization theorem $M\cong \R\x S$ in the presence of a Cauchy surface $S$. The proof of both results appeared in print only in Geroch (1970a), but it seems safe to assume that all of this must have been reasonably clear to Penrose already at the end of 1964. Indeed, the author of papers like Penrose (1963a)
 would have had no trouble
adapting the description of domains of dependence and  Cauchy surfaces in flat space in terms of characteristics in Courant \& Hilbert (1962) to curved space-times. For some history of the Cauchy problem in \GR\ see Stachel (1992), Choquet-Bruhat (2014, 2018), and  Ringstr\"{o}m (2015). } Here is a  picture from 1965 in the master's hand:
\begin{center}
\fbox{\includegraphics[width=0.8\textwidth]{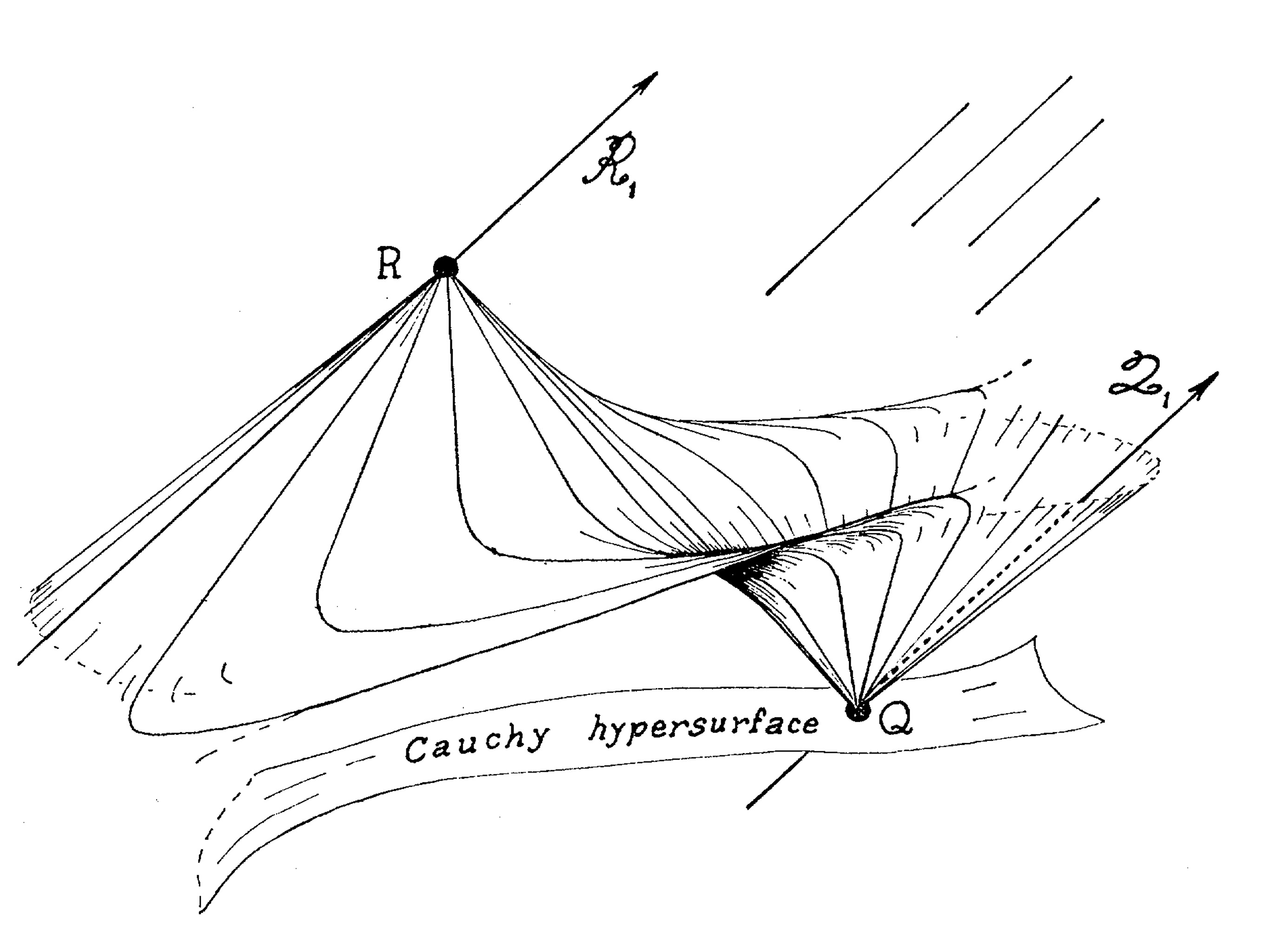}}
\end{center}
\begin{center}
\emph{Drawing of a (would-be) Cauchy (hyper)surface, taken from Penrose (1965a), p.\ 218.}
\end{center}

The concept of a trapped surface was also introduced by Penrose (1965c) himself, defined as a
\begin{small}
\begin{quote}
 closed, spacelike, two-surface $T^2$ with the property that the two systems of null geodesics which meet $T^2$ orthogonally \underline{converge} locally in future directions at $T^2$. \hfill (p.\ 58)
 \end{quote}
\end{small}
 \emph{Qua} idea and definition trapped surfaces belong to the most innovative and lasting contributions of Penrose (1965c).\footnote{See  Hawking \& Ellis (1973), chapter 9, for an early analysis of trapped surfaces in causal theory \emph{per se}, and Senovilla (2012) for a recent review. The  study of trapped surface formation from the \pde\ point of view  began with
 Schoen \& Yau (1983) and was continued by e.g.\ Christodoulou (1991, 1999a, 2009),  Dafermos (2012), Li \& Yu (2015),  and Bieri (2018).  Schoen \& Yau (1981) used the concept of a \emph{marginally outer trapped surface}
 (\textsc{mots}), which is  a Lorentzian analogue of a minimal surface in Riemannian geometry and is probably the most important variation on Penrose's trapped surface (in the latter both  expansion factors $\theta_{\pm}$ of the null congruence emanating from $T^2$ are strictly negative, whereas for a \textsc{mots} one equals zero). See Andersson, Eichmair, \& Metzger (2011) and
 Galloway, Miao, \& Schoen (2015) for reviews.  \textsc{mots}s appear in later singularity theorems with weaker assumptions (Eichmair,  Galloway, \&  Pollack, 2013)
  as well as
in the \emph{Penrose inequality}, see \S\ref{cosmic}.  }
In the presence of a radial coordinate $r$,  as in the Schwarzschild, Reissner--Nordstr\"{o}m, and Kerr solutions,
Penrose's convergence condition is equivalent to the gradient $\nabla r$ being \emph{timelike}, which in the Schwarzschild solution
happens for $r<2m$, and which in the other two (subcritical) cases is the case at least for a while after crossing the  event horizon.
The convergence condition can be stated  in terms of the null hypersurface $C$ generated by the future directed null congruence emanating from any spacelike surface $T^2$. Pick a tetrad $(e_1,e_2, L_+, L_-)$
with  $(e_1,e_2)$ spacelike and tangent to $C$, $L_{\pm}$ null and tangent (and hence orthogonal) to $C$,   such that $g(e_i,e_j)= \dl_{ij}$, $g(e_i,L_{\pm})=0$ for $i,j=1,2$,  $g(L_{\pm},L_{\pm})=0$, and $g(L_+,L_-)=-1$,
 all defined on $C$. With $\theta_{\pm}=\n_{\mu}L_{\pm}^{\mu}$, the surface
 $T^2$ is trapped iff $\theta_{\pm}<0$ throughout $T^2$ for both signs (whereas in Minkowski space-time
 one of $\theta_{\pm}$ is $>0$ whilst the other $\theta_{\mp}$ is $<0$). 
 
Thirty years later, Penrose recalled how he suddenly got the idea in the late autumn of 1964:\footnote{It is not obvious that this recollection is about trapped surfaces, but on July 2, 2022, Penrose confirmed that it was.}
\begin{small}
\begin{quote}
My conversation with  Robinson stopped momentarily as we crossed a side road, and resumed again at the other side.\footnote{This refers to  Ivor Robinson (1923--2016), as opposed to David Robinson. Both were relativists associated with Penrose via King's College London, where Penrose was a postdoc from 1961--1963 (he wrote his 1965 paper during late 1964 at Birkbeck College). [David] Robinson (2019) is a relativist's memoir that also contains interesting information about Penrose.}
 Evidently, during those few moments an idea occurred to me, but then the ensuing conversation blotted it from my mind! Later in the day, after Robinson had left, I returned to my office. I remember having an odd feeling of elation that I could not account for. I began going through in my mind all the various things that had happened to me during the day, in an attempt to find what it was that had caused this elation. After eliminating numerous inadequate possibilities, I finally brought to mind the thought that I had had while crossing the street.\footnote{It is hard to overlook the analogy with Poincar\'{e}'s famous recollection of his crucial flash of insight into the theory of Fuchsian functions, referring to an event in 1880: `At that moment I left Caen where I then lived, to take part in a geologic expedition organized by the \'{E}cole des Mines. The circumstances of the journey made me forget my mathematical work; arrived at
  Coutances we boarded an omnibus for I don't know what journey. At the moment when I put my foot on the step the idea came to me, without anything in my previous thoughts having prepared me for it; that the transformations I had made use of to define the Fuchsian functions were identical with those of non-Euclidean geometry. I did not verify  this; I did not have the time for it, since [as soon as] I sat down in the bus  I resumed the conversation already begun, but I was entirely certain at once. On  returning to Caen I verified the result at my leisure to salve my conscience.' (Grey, 2013, p.\ 217).}
  \hfill (Thorne, 1994, p.\ 227)
\end{quote}
\end{small}

The proof of the singularity theorem displays some of the unprecedented causal and topological techniques Penrose had developed in the preceding years in connection with gravitational radiation (Newman \& Penrose, 1962; Penrose, 1964, 1965b).
It is very similar to all later reformulations I am aware of:\footnote{See the second set of references in footnote \ref{refsfn}, as well as  Aretakis (2013) and Galloway (2014).}
\begin{itemize}
\item Assumptions 1, 2, 4, and 5 imply (directly) that $\p I^+(T^2)$ (which Penrose calls $B^3$) is compact;
\item Assumptions 1 and 3 imply (by contradiction) that $\p I^+(T^2)$ is non-compact.
\end{itemize}
The argument for the first point is that the null Raychaudhuri equation for $\theta_{\pm}$ (which is due to Penrose, who writes $\rho=-\half \theta_{\pm}$ for both cases) and the boundary condition $\theta_{\pm}<0$ give a focal point on each null geodesic ruling $C$.
It is left to the reader to realize that $\p I^+(T^2)=J^+(T^2)\backslash I^+(T^2)$, which follows from global hyperbolicity (which makes $J^+(T^2)$ closed), and that beyond each focal point the null geodesic enters $I^+(T^2)$ and hence leaves $\p I^+(T^2)$, in order to conclude with Penrose that 
$\p I^+(T^2)$ is compact.\footnote{See e.g.\ Aretakis (2013), paraphrased also in Landsman (2021a), \S\S6.3--6.4, for a detailed outline of this reasoning.}

Penrose's homotopical argument for the second point seems correct but has not survived;\footnote{In our discussion on July 2, 2022, he even described his proof of the second point as `stupid', and acknowledged Misner as the source of the simpler proof that later appeared in Hawking \& Ellis (1973), \S 8.2, Theorem 1.} the most rigorous proof I know of  (see e.g.\ Galloway, 2014) is obtained by combining the following  results:\footnote{The first is Lemma 3.17 in Penrose (1972); see also Minguzzi (2019), Theorem 2.87 (iii). The second is Theorem 1 in Budic, Isenberg,  Lindblom, \& Yasskin (1978); see also
Theorem 1 and Corollary 1 in Galloway (1985).}
\begin{enumerate}
\item  Any achronal boundary such as $\p I^+(T^2)$ is a (locally Lipschitz) topological hypersurface.
\item Any compact achronal topological hypersurface in a 
globally hyperbolic space-time  is a Cauchy surface; hence by Geroch (1970a), Property 7, any Cauchy surface in this space-time is compact.
\end{enumerate}
\section{Intermezzo: Singularities revisited and refined}\label{intermezzo} 
 In his PhD Thesis, Hawking  writes without any further ado (or mention of inextendibility):
 \begin{small}
\begin{quote}
any model must have a singularity, that is, it cannot be a geodesically complete $C^1$, piecewise $C^2$ manifold. \hfill  (Hawking, 1965, \S 4.1)
\end{quote}
\end{small}
 And similarly in the paper announcing his singularity theorem,  citing Misner (1963) at  this point:
  \begin{small}
\begin{quote}
Space-time is said to be singularity free if it is time-complete (all timelike geodesics can be extended to arbitrary length) and if the metric is a $C^2$ tensor field. \hfill ( Hawking, 1966a, p.\ 444) 
\end{quote}
\end{small}
In his subsequent  Adams Prize Essay,\footnote{See Ellis (2014) and \url{https://en.wikipedia.org/wiki/Adams_Prize} for the history of this award. `The Adams Prize is awarded jointly each year by the Faculty of Mathematics and St John’s College to UK-based researchers, under the age of 40, doing first class international research in the Mathematical Sciences.' Each year a topic is set, which in 1966 was \emph{Geometric problems of relativity}. Although Penrose won the prize, Hawking was awarded an `auxiliary Adams Prize' in the same year. The memoir of Ellis (2014) about Hawking is especially valuable since he was around himself at the time: `After the publication of [Penrose's]  paper in January 1965, the members of Dennis
Sciama's general relativity group in the Department of Applied Mathematics and
Theoretical Physics at Cambridge University (particularly Stephen Hawking, myself,
and Brandon Carter) hurriedly tried to learn the new methods that Penrose
had introduced. We were assisted in this by discussions with Felix Pirani and the
group at King's College, London; with John Wheeler and Charles Misner, who visited
Cambridge from the USA for an extended period; and with Roger Penrose and Bob
Geroch, who was visiting Penrose at Birkbeck College, London. In particular we had a
one day seminar in Cambridge attended by the members of the King's College group,
where I and Brandon Carter summarized our understandings of the ingredients of
Penrose's theorem. (\ldots) Stephen arrived at [his] results by discussions
with the Cambridge group that under Dennis Sciama's guidance met to discuss
ideas at tea time each day, and with the London groups (as well as attending many
seminars, we used to regularly catch the train to attend lectures on general relativity
at King's College, London).' (Ellis, 2014, pp.\ 3--4). See also Tipler, Clarke, \& Ellis (1980), pp.\ 135--136.
}
 Hawking explicitly assumes inextendibility and proposes to:
\begin{quote}
\begin{small}
take timelike and lightlike geodesic incompleteness as our definition of a singularity of space-time. \hfill
\\ \mbox{} \hfill (Hawking, 1966b, \S 6.1)
\end{small}
\end{quote}
All of this was directly inspired by  Penrose (1965c), but it seems to have been Hawking, rather than Penrose, who first advocated causal geodesic incompleteness as the \emph{definition} of a space-time singularity. 
In a later defense of this definition, Hawking and Ellis make both a physical and a pragmatic point:\footnote{Also cf.\ Geroch (1968), p.\ 534:  `Finally, we remark that geodesic incompleteness (\ldots) is commonly used as a definition of a singularity because with such a definition one can show that large classes of solutions of Einstein's equations are singular.'}
 \begin{quote}\begin{small}
Timelike geodesic incompleteness has an immediate physical significance in that it presents the possibility that there could be freely moving observers or particles whose histories did not exist after (or before) a finite interval of proper time. This would appear to be an even more objectionable feature than infinite curvature and so it seems appropriate to regard such a space as singular. (\ldots)  The advantage of taking timelike and/or null incompleteness as being indicative of the presence of a singularity is [also] that on this basis one can establish a number of theorems about their occurrence. \\ \mbox{} \hfill  (Hawking \& Ellis, 1973, p.\ 258)
 \end{small}
\end{quote}
As we have seen, crucial nuance is lost in doing so. Although Geroch (1966) uses practically the same definition as Hawking, also citing Misner (1963), his subsequent analysis (Geroch, 1968) is more critical:
 \begin{quote}
\begin{small}
\begin{enumerate}
\item[(a)] there is no widely accepted definition of a singularity in general relativity;
\item[(b)] each of the proposed definitions is subject to some inadequacy. \hfill   (Geroch, 1968, p.\ 526)
\end{enumerate}
\end{small}
\end{quote}
This paper  is still worth reading, as is its more technical successor Geroch (1970b); see also Clarke (1993), Earman (1995), Beem,   Ehrlich, \& Easley (1996), Senovilla (1998), and Curiel (1999, 2019a). 

Almost 60 years later, none of the problems that arose from identifying space-time singularities with incomplete causal geodesics has been resolved:  timelike, null (i.e.\ lightlike), causal, and general geodesic incompleteness are all inequivalent to each other;  each is in turn inequivalent to incompleteness with respect to curves with either arbitrary or bounded acceleration; geodesically incomplete manifolds (of any kind) exist without  curvature singularities; global hyperbolicity is compatible with geodesic incompleteness; 
geodesic completeness is compatible with a lack of  global hyperbolicity; \emph{et cetera}. 
\newpage

 Penrose (1974) gave a new definition of a singularity (i.e.\ Definition \ref{CSdef1} below)
`which is not quite the same as that suggested by the singularity theorems' (p.\ 85). 
It has two novel and distinguishing features:
\begin{enumerate}
\item It is \emph{perspectival}, in that not just the singularity matters but also its relationship to  points which it can influence. This is crucial for understanding Penrose's mature idea of cosmic censorship (\S\ref{cosmic}).
\item \emph{Inextendible timelike curves} replace
 \emph{incomplete causal  geodesics} as marks of singularities.
\end{enumerate}
The change from `causal' to `timelike' is innocent,\footnote{Cf.\ Theorem (2.3) in  Geroch, Kronheimer, \&  Penrose (1972);  one could even use lightlike curves (Flores \& S\'{a}nchez, 2008,  \S 3.3). 
 Causal \emph{geodesics} would   lead to some weaker causality condition than global hyperbolicity, cf.\ Theorem \ref{P79theorem} below.} but the change from \emph{incomplete geodesics} to  \emph{inextendible curves}  is not (this change is required by the proof of Theorem \ref{P79theorem}  below).\footnote{It is the second (`converse') part of the proof of Theorem \ref{P79theorem} below that does not work for causal  geodesics instead of  curves, since the curve $c$ constructed there is not necessarily a geodesic. The fact that we just get a causal curve at that point goes back to the definition of domains of dependence and  Cauchy surfaces in terms of causal curves rather than geodesics.} 
 The key difference lies in the possibility that 
 an inextendible causal curve may go off to infinity and yet count as a singularity in the new sense, i.e.\  as a   `locally naked' one.\footnote{Inextendible causal curves may also hover around in a compact set, but this is impossible in strongly causal space-times.}
  For example, even $M=\R^4$ carrying plane gravitational waves, which is geodesically complete and strongly causal but not globally hyperbolic, will now count as singular,\footnote{This example, which was kindly suggested to me in this context  by Senovilla, is even due to Penrose (1965a)!  A metric $g=dudv-dy^2-dz^2+H(u,y,z)du^2$
   on $\R^4$, which describes gravitational or electromagnetic waves,
  is Ricci-flat iff $(\p_y^2+\p_z^2)H=0$, which is solved by
   $g=dudv-dy^2-dz^2+f(u)(y^2-z^2)+g(u)yz$. See  Beem, Ehrlich, \& Easley (1996), Chapter 13, Flores \&  S\'{a}nchez (2003, 2008), and  Garc\'{\i}a-Parrado \&  Senovilla  (2005) for details and generalizations. If $f\neq 0$ such  space-times cannot be asymptotically flat at null infinity (cf.\  footnote \ref{ccfn}), which would therefore be desirable. Also cf. Penrose (1979), p.\ 623.} as does anti-de Sitter.\footnote{In this context, anti-de Sitter should be seen as a special case of the Malament--Hogarth space-times (Hogarth, 1996, Manchak, 2020), which are 
  never globally hyperbolic and yet typically geodesically complete (as indeed is anti-de Sitter).}
Consequently,  Definition \ref{CSdef1} accepts more singularities than the singularity theorems, and \emph{excluding} these new singularities 
therefore certainly excludes all old-school singularities: it will be an \emph{unnecessarily strong} condition.
However, in \emph{strongly causal space-times that are asymptotically flat at null infinity}  the right intuition to think about the ``new'' singularities remains: causal curves that either crash into a curvature singularity or reach the edge of an extendible space-time.\footnote{Penrose (1974, 1979)  assumes that $(M,g)$ is strongly causal, which on the causal ladder is one step above being past distinguishing (Minguzzi, 2019, chapter 4,  Definition 4.46), i.e.\ the  condition that suffices to make sense of Definition \ref{CSdef1}. However, strong causality is necessary for Theorem \ref{P79theorem} below.
This is true for Penrose's original proof because of his reliance on the theory of TIPs, and in my version of the proof this is because of the invocation of Theorem 2.53 in Minguzzi (2019), where the case distinction I need
relies on strong causality via his Remark 2.54 and Theorem 2.80.  }

 Penrose states his new definition in the language of terminal indecomposable sets (Geroch, Kronheimer, and Penrose, 1972), but this language can be avoided. A simpler account is as follows:
 \begin{enumerate}
\item  Suppose that  $(M,g)$ is \emph{past distinguishing},\ i.e.,  $I^-(x)=I^-(y)$ implies $x=y$.
This property allows us to exchange properties of points $x$ for properties of their timelike pasts $I^-(x)$.\footnote{\label{wcsdoc} One may also use  \emph{past-directed} inextendible causal curves, replacing $I^-(\cdot)$  by  $I^+(\cdot)$, etc. For strong cosmic censorship this definition is  equivalent to the given one, as follows from Theorem \ref{P79theorem} below, which follows either way (Penrose, 1979).} 
\item By definition, if $z\in I^-(x)$ then $z$ can signal to $x$, or influence $x$; we may say that $z$ is \emph{naked} for $x$. 
For the next step, we note that  $z\in I^-(x)$ is equivalent to the property $I^-(z)\subset I^-(x)$.
\item If $z$ is the endpoint of some future-directed timelike curve $c$, then  $I^-(z)\subset I^-(x)$ if and only if  
 \beq
I^-(c)\subset I^-(x). \label{Isubset}
\eeq
\item Crucially, this inclusion is also defined if $c$ has no endpoint (i.e.\ is inextendible). This suggests:
\end{enumerate}
 \begin{definition}\label{CSdef1}
 A future-directed future-inextendible causal curve  $c$ in $M$ is \emph{naked} 
for $x\in M$ if \er{Isubset} holds.
   \end{definition}
 For example,  in Minkowski space-time, for fixed $x$  take $z\in I^-(x)$ and remove $z$. 
 Then any  fd future-inextendible timelike curve $c$ whose endpoint would have been $z$ satisfies \er{Isubset}. 
 \section{Cosmic censorship: Penrose}\label{cosmic}
 Let us now return to Penrose (1965c). In \S\ref{exegesis} I already noted that his emphasis on the inextendibility (or, as he put it, completeness) of the underlying space-time shows that he was fully aware that (null) geodesic incompleteness alone was insufficient to point to a curvature (or ``physical'') singularity, i.e., a ``hole''.  
Similarly, Penrose realized that the ``black'' in our (later) term ``black hole'' was as yet lacking:\footnote{The rise of the term ``black hole'' is discussed in the appendix to Landsman (2021b), which was written by Erik Curiel.
See also Israel (1987) and Thorne (1994). I will freely use this material in what follows. Briefly, although it seems that the term was coined by Dicke and/or Wheeler, Penrose (1969) marked its first use in the scientific literature, where for the first few years it was usually surrounded by scare quotes. Other pioneers of the term were Bardeen, Carter, Christodoulou, and Hawking.
}
 \begin{quote}
\begin{small}
Whether or not ``visible'' singularities inevitably arise under appropriate circumstances is an intriguing question not covered by the present discussion. \hspace{45pt} (Penrose, 1965c, footnote 9, p.\ 59)
\end{small}
\end{quote}
This was taken up in his magisterial and very influential paper `Gravitational collapse: The role of general relativity' (Penrose 1969), in which Penrose introduced what we now call  \emph{weak cosmic censorship}:\footnote{For reviews of cosmic censorship before the \pde\ people took over see Geroch \& Horowitz (1979), \S 5.4, 
Tipler, Clarke, \& Ellis (1980), \S 6, Earman (1995), Chapter 3, and Joshi (2007), Chapter 4. See \S\ref{future} for the \pde\ approach.}
    \begin{quote}
\begin{small}
We are thus presented with what is perhaps the most fundamental question of general-relativistic collapse theory, namely: does there exist a ``cosmic censor'' who forbids the appearance of naked singularities, clothing each one in an absolute event horizon? In one sense, a  ``cosmic censor'' can be shown \emph{not} to exist. For it follows from a theorem of Hawking that the ``big bang'' singularity is, in principle, observable. But it is not known whether singularities observable from outside will ever arise in a generic \emph{collapse} which starts off from a perfectly reasonable nonsingular initial state. \\ \mbox{} \hfill (Penrose, 1969,  p.\ 1162)
\end{small}
\end{quote}

For a mathematician like Penrose, this conjecture (here phrased as a question) relied on a precise definition of an ``absolute event horizon''. Like his 1965 singularity theorem, also this concept used mathematical ideas that Penrose had initially developed for gravitational waves, especially his notion of null infinity $\CI$, pronounced ``scri'' (Penrose, 1963b, 1964, 1968). Perhaps because Penrose (1969) is mainly addressed to physicists, he relegates
this pretty important matter to a footnote, which reads:\footnote{\label{ccfn} Instead of Penrose's weakly asymptotically simple space-times, I use the cleaner notion of space-times $(M,g)$ that are \emph{asymptotically flat at null infinity} (Chru\'{s}ciel, 2020, \S 3.1). 
This means that $(M,g)$ has a
conformal completion $(\hat{M},\hat{g})$, i.e., a conformal embedding $\io:M\hookrightarrow\hat{M}$ such that $\p\hat{M}=\hat{M}\backslash \io(M)\equiv
\CI=\CI^+\cup\CI^-$, where $\CI^{\pm}:=\p\hat{M}\cap J^{\pm}(M)$. Writing $\io^*\hat{g}=\io^*\Om^2g$ for some smooth function $\Om:\hat{M}\raw[0,\infty)$, one asks
$\Om>0$ on $\iota(M)$, $\Om=0$ on $\p \hat{M}$, and $d\Om\neq 0$ on $\p \hat{M}$.
Adding fall-off conditions on the Ricci tensor guaranteeing that $\CI^{\pm}$ are null hypersurfaces in $\hat{M}$, one finally requires that
$\CI^{\pm}\cong\R\x S^2$.  To simplify the notation I always assume that null infinity $\CI$ is such that its timelike past $I^-(\CI^+)$ lies in $M$, i.e.\ does not intersect the boundary $\p\hat{M}$. If this is not the case (as e.g. in anti-de Sitter space),  just write $I^-(\CI^+)\cap M$ instead of $I^-(\CI^+)$ everywhere.}
  \begin{quote}
\begin{small}
In a general space-time with a well-defined external future infinity, the absolute event horizon would
be defined as the boundary of the union of all timelike curves which escape to this external future
infinity. In the terminology of Penrose (1968), if $M$ is a weakly asymptotically simple space-time, for
example, then the absolute event horizon in $M$ is $\dot{I}_-(\CI^+)$ [$=\p I^-(\CI^+)$]. \\ \mbox{} \hfill (Penrose 1969, footnote 3, p.\ 1146 of the 2002 reprint)
\end{small}
\end{quote}
 Penrose leaves it to the reader to extract the mathematical definition of a black hole from this; and
as in the case of the definition of a singularity, it was again Hawking who was subsequently more explicit:
  \begin{quote}
\begin{small}
A black hole on a spacelike surface [$S$] is defined to be a connected component of the region of the surface bounded by the event horizon [i.e.\ $M\backslash I^-(\CI^+)\cap S$].\hfill  (Hawking 1972, Abstract, p.\ 152)
\end{small}
\end{quote}
Later (p.\ 156) he explains that this is a region `from which there is no escape to $\CI^+$'. 
 This definition is repeated in Hawking \& Ellis (1973), \S 9.2, and is still standard,\footnote{\label{Cur} This does not mean it is uncontroversial (Curiel, 2019b). The opposition between Rovelli and Wald is especially telling. Rovelli states that Penrose's definition 
 of an absolute event horizon is `very useless, because it assumes we can compute the future of real black holes, and we cannot' (Curiel, 2019b, p.\ 30), to which Wald replies that
`I have no idea why there should be any controversy of any kind about the definition of a black hole. There is a precise, clear definition in the context of asymptotically flat spacetimes (\ldots) I don't see this as any different than what occurs everywhere else in physics, where one can give precise definitions for idealized cases but these are not achievable/measurable in the real world.' (\emph{ibid.}, p.\ 32).}
 usually in the simple form $M\backslash I^-(\CI^+)$.
 
\noindent However, since there can be holes (in the sense of curvature singularities) that are not black in having no event horizons,\footnote{Simply take the Schwarzschild metric for $m<0$, or the supercharged Reissner--Nordstr\"{o}m metric (i.e.\ $e^2>m^2>0$), or the fast rotating Kerr solution  (i.e.\ $a^2>m^2>0$); these are of course the simplest examples of Penrose's naked singularities.}  and there may be event horizons that do not cover any kind of singularity,\footnote{For an artificial  example,  in the usual conformal completion of Minkowski space-time $\Mi$ (Penrose, 1964),
truncate $\CI^+=\{(p,q)\mid p=\pi/2, q\in (-\pi/2,\pi/2)\}$ to $\{(p,q)\mid p=\pi/2, q\in (-\pi/2,0)\}$. This  turns the region $J^+(0)$ into  a fake black hole in flat space-time. What I call black \emph{objects} also include 
regular (i.e.\ geodesically complete)  black ``holes''. See e.g.\ Frolov (2016),  Spallucci \& Smailagic (2017),  Carballo-Rubio \emph{et al.} (2020), and Simpson \& Visser (2022) and references therein
 \label{Heino}
} it might be better to call (a connected component of) $M\backslash I^-(\CI^+)$ a \emph{black object} instead of a black hole. 
In any case, his definition of an event horizon led Penrose  to a more precise definition of (weak) cosmic censorship:
  \begin{quote}
\begin{small}I first review what is now the ``standard'' picture of gravitational collapse to a
black hole---which I shall refer to as the \emph{establishment} viewpoint. The picture is, in
fact, a very good one, with a remarkable degree of internal consistency. But it is
worth emphasizing that, among other things, this picture does depend on a very
big assumption, that of \emph{cosmic censorship}. In my own view, this assumption might
well be false in suitable circumstances. I believe that the consequences of possible
violations of cosmic censorship are certainly worth considering seriously (\ldots)
In effect, this principle states that naked singularities do not develop
out of an initially nonsingular state (\ldots). A naked singularity
may be said to exist in a space-time $M$ if there is a well-defined external
infinity (that is, $\CI^+$, if $M$ is weakly asymptotically simple) from which timelike
curves may be drawn into the past that terminate on the singularity. I
shall not make precise, here, the concept of ``terminating on a singularity'' but refer
instead to Hawking's concept of \emph{asymptotic predictability}.\footnote{A space-time 
$(M,g)$ is (future) asymptotically predictable  from a partial Cauchy surface $S$ if
$\CI^+\subset\overline{D^+(S)}$, where the domain of dependence $D^+(S)$ is computed in
  the conformal completion $\hat{M}$ of $M$ (Hawking \& Ellis, 1973, p.\ 310), cf.\ footnote \ref{ccfn}.  A violation of this condition would come from some null geodesic outside $D^+(S)$, presumably marking a singularity at least in the spirit of Penrose (1965c), reaching $\CI^+$. This would makes such a singularity visible to asymptotic observers.}  This is a precise condition
on a space-time $M$ that may be taken as a statement that $M$ accords with a
form of cosmic censorship. Imposing such a condition, we deduce the existence of an
absolute event horizon $E$, which encloses but does not intersect the singular region.
This horizon $E$ may be defined as the boundary of the past of the external infinity. \hfill (Penrose, 1973, p.\ 125)
\end{small}
\end{quote}
The first part is tongue in cheek, since the ``establishment''  included---or even \emph{was}---Penrose himself!
The `form of cosmic censorship' probably refers to Hawking's (1972) theorem to the effect that  any future trapped surface lies entirely within a black hole.\footnote{See also Proposition 9.2.1 in Hawking \& Ellis (1973), corrected by Claudel (2000).} Penrose (1973) goes on to develop the \emph{Penrose inequality}, which he saw as a test for cosmic censorhip he expected to fail (which it has not, so far).\footnote{See Bray \&  Chru\'{s}ciel  (2004) and Mars (2009)  for reviews of the Penrose inequality. See also  Landsman (2021a), \S 10.11 for a brief summary. A special case, the \emph{Riemannian Penrose inquality}, has been proved; see Lee (2019) for details.}

The challenge of making  `the concept of ``terminating on a singularity''' more precise was taken up in Penrose (1974), see also Penrose (1979), and led to Definition \ref{CSdef1} above. A definition of cosmic censorship should then specify the nature of the points $x$ relative to which ``singular'' curves $c$ are naked. Hence it would have been natural if Penrose had proposed e.g.\  the following definition (cf.\ footnote \ref{ccfn}):\footnote{Definition \ref{dwcc} is  stated in words by Geroch \& Horowitz (1979), pp.\ 274--277.
 See also Earman (1995), pp.\ 74--78.}
 \begin{definition}\label{dwcc}
Let $(M,g)$ be a strongly causal space-time that is asymptotically flat at null infinity.
\begin{enumerate}
\item An \emph{$N$-naked singularity} in $M$ with respect to some region of exposure $N\subset\hat{M}$ is a future-inextendible causal curve  $c$ in $M$ that is naked for some point $x\in N$, i.e., the pair $(c,x)$ satisfies $I^-(c)\subset I^-(x)$.
\item The  \emph{$N$-weak cosmic censorship conjecture} states that a ``physically reasonable'' space-time $(M,g)$ 
contains no $N$-naked singularities, i.e.,
no curve $c$ in $M$ \emph{and} point $x\in N$ as specified in 1.\ exist.
\end{enumerate}
\end{definition}
Paraphrasing Geroch \& Horowitz (1979), p.\ 274, the set $N$ represents the region of space-time (or possibly of its conformal completion $\hat{M}$) `in which observers could detect that their universe is singular.' 
This definition is quite subtle and excludes for example that  the big bang or the big crunch are seen as naked singularities (and the ensuing cosmos as violating weak cosmic censorship): for the big bang, which we  \emph{can} ``see'', no such curve $c$ exists (any relevant  $c$ is \emph{past} inextendible), whereas for the big crunch no $x$ would exist.
See Geroch \& Horowitz (1979), pp.\ 274--277 and Penrose (1979), \S 12.3.2.\footnote{In Penrose (1979), p.\ 618, a `physically reasonable' space-time is taken to mean: `A system which evolves, according to classical general relativity with reasonable equation of state, from generic non-singular initial data on a suitable Cauchy surface.' But this is stated in connection with \emph{weak} cosmic censorship. In strong cosmic censorship, the above formulation is problematic since it suggests that
the `evolved' space-times in question are globally hyperbolic, which by Theorem \ref{P79theorem} makes 
strong cosmic censorship trivially true. I return to this problem in \S\ref{future} below. 
Penrose (1974) does not state genericity conditions.}

Various choices of the region $N$  have (often implicitly) been proposed in the literature, for example:\vspace{-1mm}
\begin{itemize}
\item $N=I^-(\CI^+)$, the complement of the black hole region in $M$ (my interpretation of Penrose, 1969);
\item  $N=I^-(\CI^+)\cap I^+(\CI^-)\equiv\mathcal{D}(M,\CI)$, i.e.\ the domain of outer communication(s) in $M$;\footnote{See e.g.\ 
 Chru\'{s}ciel (2020), \S 3.1, for this notion. Modern black hole uniqueness theorems like  Theorem 3.1 in Chru\'{s}ciel,  Lopes Costa, \& Heusler (2012) assume weak cosmic censorship \emph{defined} as global hyperbolicity of $I^-(\CI^+)\cap I^+(\CI^-)$. }  
 \item  $N=\mathcal{D}(M,\CI)\cup\CI$ with $\CI=\CI^+\cup\CI^-$ (Penrose, Sorkin, \& Woolgar, 1993);
 \item $N=J^-(\CI^+)\cap J^+(S)$, where $S$ is some partial Cauchy surface in $M$.\footnote{This comes tantalizingly close to Hawking's future asymptotically predictability but is not equivalent to it, see 
Kr\'{o}lak (1986), especially Lemma 2.10. See also Wald (1984), \S 12.1 and Chru\'{s}ciel (2020), \S 3.5.1.}
\item $N= J^-(\CI^+)$ (Tipler, Clarke \& Ellis, 1980);\footnote{In fact, Tipler, Clarke \& Ellis (1980), p.\ 176, \emph{define} weak cosmic censorship as   global hyperbolicity  of $J^-(\CI^+)$.}
\item $N=\CI^+$  (Geroch \& Horowitz, 1979);\footnote{Their definition of (weak) cosmic censorship is that 
the closure of $N$ (defined in $M$) in $\hat{M}$ does not intersect $\CI^+$ (p.\ 280).}
\item $N=M$ (Penrose, 1974, 1979), see below (strong cosmic censorship).\footnote{Since and $\hat{M}$ and $\CI^+$ play no role, this is of course defined without any need for asymptotic flatness in Definition \ref{dwcc}.}\vspace{-1mm}
\end{itemize}
The first and last choices seem optimal in having a clean characterization via Theorems \ref{twcc} and \ref{P79theorem}. First:
\begin{theorem}\label{twcc}
Let $(M,g)$ be a strongly causal space-time that is asymptotically flat at null infinity.
Then
$(M,g)$ contains no naked singularities with respect to  $I^-(\CI^+)$ iff $I^-(\CI^+)$ is globally hyperbolic.
\end{theorem}
This is proved in the same way as Penrose's Theorem \ref{P79theorem} below, adding a case distinction $x\in I^-(\CI^+)$ and $x\notin I^-(\CI^+)$ that does not upset the result. 
If one accepts the choice $N=I^-(\CI^+)$, Theorem \ref{twcc} might of course serve to \emph{define} weak cosmic censorship as  global hyperbolicity of $I^-(\CI^+)$. Except for $N=M$ there seems to be no complete analogue of Theorem \ref{twcc} for the other choices of $N$; at best,
there is a one-sided implication to global hyperbolicity of $N$. In any case, I side with Earman (private communication) that one may regard weak cosmic censorship as a \emph{family} of conjectures, one for each relevant region $N$.

But Definition \ref{dwcc} is ahistorical! Instead, Penrose (1974, 1979) immediately moved on,\vspace{-1mm}
arguing that:\footnote{Geroch \& Horowitz (1979) do not endorse this: `One would not, however, wish to regard [Kerr space-time with $a^2<m^2$] as a counterexample to cosmic censorship, for it is only the ``nearby'' observers who can detect the singular behavior, not the ``distant'' observers.' (p.\ 277). Slowly rotating Kerr space-time is not globally hyperbolic but it does have an event horizon.}
  \begin{quote}
\begin{small}
It seems to me to be comparatively unimportant whether the observer himself can escape to infinity. Classical general relativity is a scale-invariant theory, so if locally naked singularities occur on a very tiny scale, they should also, in principle, occur on a very large scale in which a `trapped' observer could have days or even years to ponder upon the implications of the uncertainties introduced by the observations of such a singularity. (\ldots) Indeed, for inhabitants of recollapsing closed universes (as possibly we ourselves are) there is no `infinity', so the question of being locally `trapped' is one of degree rather than principle.  
It would seem, therefore, that if cosmic censorship is a principle of Nature, it should be formulated in such a way as to preclude such \emph{locally} naked singularities. \\ \mbox{}\hfill (Penrose, 1979,  p.\ 619)
\end{small}
\end{quote} 
Thus `locally naked singularities' are defined via Definition \ref{dwcc} by choosing
  $N=M$.
Then the key result (which \emph{a posteriori} justifies his use of curves instead of geodesics) is due to Penrose (1974, 1979):
   \begin{theorem}\label{P79theorem} 
 A strongly causal space-time has no locally naked singularities iff it is globally hyperbolic.
\end{theorem}
\emph{Proof.}\footnote{This proof is taken from Landsman (2021a), \S 10.4. Penrose (1979) gives a different argument using TIPs.} 
I first prove the implication  ``global hyperbolicity $\Raw$ no locally naked singularities'' by contradiction. Suppose that $(M,g)$ is globally hyperbolic and  that \er{Isubset} holds for some $c$ and $x$.
 Take $y$ on $c$ and then take a future-directed sequence $(y_n)$ of points on $c$, with $y_0=y$. 
 Because of \er{Isubset} this sequence lies in  $J^+(y)\cap J^-(x)$, which is compact by assumption. Hence 
 $(y_n)$ has a limit point $z$ in  $J^+(y)\cap J^-(x)$. 
 
 Now define curves $(\gm_n)$ as the segments of $c$ from $y$ to $y_n$. By Theorem 2.53  in Minguzzi (2019), case (i), which is a sharp version of the limit curve lemma of causal theory going back to Penrose (1972),
 these curves have a uniform limit $\gm$.  Its arc length (as measured by an auxiliary complete Riemannian metric)
  is, on the one hand, infinite (since $c$ is endless and hence has infinite arc length, which is approached as the $y_n$ move up along $c$). But on the other hand  it is finite, since $\gm$  ends at $z$ (and fd continuous causal curves have finite arc length iff they have an endpoint). 
   Hence \er{Isubset} cannot be true for any $c$ and $x$.

The (contrapositive) proof of the converse implication relies on the following lemma:\footnote{Here I do follow Penrose (1979), p.\ 624. The lemma combines Propositions 5.20 and 5.5 (h) in Penrose (1972). }
\begin{lemma}\label{P72.5.20}
Let $(M,g)$ be a space-time, let $S\subset M$ be closed and achronal, and let  $x,y\in M$.
\begin{enumerate}
\item If $y\in\mathrm{int}(D^-(S))$, then $J^+(y)\cap J^-(S)$ is compact. In particular, taking
$S=\p I^-(x)$\\ and assuming  $y\in I^-(x)$, it follows that $J^+(y)\cap J^-(x)$ is compact.
\item We have $\mathrm{int}(D^-(S))=I^-(S)\cap I^+(D^-(S))$.
\end{enumerate}
\end{lemma}\noi
 To prove the converse direction  
 of Theorem \ref{P79theorem}, assume that
$(M,g)$ is not globally hyperbolic.
 Then, under the assumption of strong causality,
there are $x,y$ for which $J^-(x)\cap J^+(y)$ is not compact.\footnote{Unless $M$ is compact this is true even without the assumption of
strong causality, see Hounnonkpe \& Minguzzi (2019).}
 We may assume that $y\in I^-(x)$. Part 1 of Lemma \ref{P72.5.20} gives 
$y\notin \mathrm{int}(D^-(\p I^-(x)))$. Part 2 gives some $y'\in  I^-(x)$ with $y'\notin D^-(\p I^-(x))$, so that, by definition of $D^-$,  there exists some fd future-inextendible curve $c$ from $y'$ that avoids $\p I^-(x)$. Since 
 $y'\in  I^-(x)$, this curve does lie in $I^-(x)$, and hence \er{Isubset} holds.\footnote{All this can be checked in the Minkowskian example following Definition \ref{CSdef1}, where, assuming $z\in I^+(y)$,  the removal of $z$ ruins compactness of  $J^+(y)\cap J^-(x)$ and hence global hyperbolicity\index{global hyperbolicity}. The existence of $c$ is trivial. }   \enp 
\begin{definition}\label{tlns}
The  \emph{strong cosmic censorship conjecture} states that no ``physically reasonable''  space-time $(M,g)$ 
contains locally naked singularities, i.e.\ pairs $(c,x)$ as in Definition \ref{CSdef1} such that $I^-(c)\subset I^-(x)$. 
Equivalently, the conjecture states that any  ``physically reasonable'' space-time is globally hyperbolic.
\end{definition}

 \smallskip
 
\noindent 
 In this formulation, strong cosmic censorship ($N=M$) implies weak cosmic censorship ($N=I^-(\CI^+)$).\footnote{This follows from the definitions, but it also follows from Theorems \ref{twcc} and \ref{P79theorem} and the observation that if $(M,g)$ is globally hyperbolic, then so is  $I^-(\CI^+)$: if $x\in J^+(y)$ for $x,y\in I^-(\CI^+)\subset M$, then $J^+(y)\cap J^-(x)\subset 
   I^-(\CI^+)$.}  
   
   Under the assumptions of Penrose's singularity theorem 
one can  prove an analogue of Hawking's (1972) theorem on the invisibility of trapped surfaces (cf.\ Hawking \& Ellis, 1973, Proposition 9.2.1):
   \begin{theorem}\label{HTST}
   Under the assumptions of Theorem \ref{P65theorem} and the additional clause that $(M,g)$ be asymptotically flat at null infinity,  any  future trapped surface must lie entirely within the ``black region'' $M\backslash I^-(\CI^+)$.
\end{theorem} 
\noi I just sketch the proof, as the details are similar to Hawking's.
 If the trapped surface $T^2$ were to (partly) lie in  $I^-(\CI^+)$, then also part of $\p I^+(T^2)$ lies in  $I^-(\CI^+)$. Hence some of the lightlike geodesics $\gm$ ruling $\p I^+(T^2)$ with past endpoint on $T^2$ would reach $\CI^+$ and hence have infinite length. But the definition of a trapped surface excludes this, as in   the proof of Theorem \ref{P65theorem}, since it forces each $\gm$ to be incomplete. \enp
\smallskip

\noi Since this forces $M\backslash I^-(\CI^+)$ to be non-empty, there must  be an event horizon $\p  I^-(\CI^+)$, which by the same argument must cover not only any trapped surface but also all ensuing incomplete null geodesics.
 \section{Cosmic censorship: From Penrose to \pde}\label{future}
 Theorem \ref{HTST} bridges part of the gap between Penrose's 1965 singularity theorem and its intended application to black holes, in that in the asymptotically flat setting an event horizon comes for free. 
 However, according to this analysis (whose centerpiece is Penrose's Theorem \ref{P79theorem}), strong cosmic censorship (\`{a} la Penrose, cf.\  Definition \ref{tlns})
 is in fact one of the assumptions of the theorem, so that its implication of weak cosmic censorship (once again in the sense of Penrose, i.e.\ defined by taking $N=I^-(\CI^+)$ in Definition \ref{dwcc}) is hardly surprising. And we are also left with the issue of the (in)extendibility of space-time. 
 
 With hindsight, it would have been natural if, one the basis of his (1965c), Penrose had defined:\footnote{Compare Tipler, Clarke, \& Ellis (1980), p.\ 167: `The \emph{strong cosmic censorship principle} is said to hold if the \emph{entire} space-time $(M,g)$ is globally hyperbolic, i.e., if there are no nontrivial Cauchy horizons, and the  \emph{weak cosmic censorship principle} is said to hold if all breakdowns of global hyperbolicity occur inside a black hole, i.e.\ if the region $J^-(\CI^+)$ is globally hyperbolic.'}
  \begin{itemize}
\item \emph{weak cosmic censorship} as the existence and stability of event horizons (in case of singularities);
\item \emph{strong cosmic censorship} as the nonexistence or instability 
of Cauchy horizons (in all cases).
\end{itemize}
This would have directly addressed the above gap. In the first (weak) case this is indeed what he did, but he
subsequently  replaced this by the version of strong cosmic censorship that really interested him.\footnote{Penrose (1999) still states that the inference from geodesic incompleteness to a black hole requires weak cosmic censorship in a version that looks like a form of Definition \ref{dwcc} with either $N=I^-(\CI^+)$ or $N=\CI^+$. But, as in Penrose (1979),  he then 
moves on to strong cosmic censorship (in his version, of course) as the version one really wants.
Note that (his) strong cosmic censorship does not imply the $N=\CI^+$-weak version! In our discussion on July 2, 2022, Penrose said that for him the essence of strong cosmic censorship had always been the avoidance of timelike singularities, and that (in)extendibility was  secondary.
} However, via a detour explained below, the latter eventually did lead to the second bullet point.

The current versions of both cosmic censorship conjectures are based on the  initial-value approach or \pde\ approach to \GR, whose starting point may be taken to be the celebrated theorem of Choquet-Bruhat \& Geroch (1969),\footnote{The original source is Choquet-Bruhat \& Geroch (1969), who merely sketched a proof (based on Zorn's lemma). Even the 800-page textbook by  Choquet-Bruhat (2009)
does not contain a proof of the theorem (which is Theorem XII.12.2); the treatment in Hawking \& Ellis (1973), \S 7.6,  is slightly more detailed but far from complete, too. 
 Ringstr\"{o}m (2009) is a book-length exposition, but ironically the proof of Theorem 16.6
 is wrong; it  is corrected in Ringstr\"{o}m (2013), \S 23. A constructive proof was given by Sbierski (2016).
For a summary see Landsman (2021a), \S 7.6, which also contains further information.}
which states that for each initial data triple  $(S, h,K)$ satisfying the vacuum constraints, where $(S,h)$ is a complete $3d$ Riemannian manifold carrying an additional symmetric 2-tensor 
 $K$, there exists a maximal globally hyperbolic development or \mghd\ $(M,g,i)$, where $(M,g)$ is a space-time  satisfying the vacuum Einstein equations and $i:S\hookrightarrow M$ is an embedding  such that $g$ returns the initial data $(h,K)$ on $i(S)$, in that $i^*g=h$ and $K$ is the extrinsic curvature of $i(S)\subset M$. 
 The space-time $(M,g)$ contains
$i(S)$ as a Cauchy surface and hence is globally hyperbolic (as the name \mghd\ suggests).
 Moreover, the triple $(M,g,i)$ is unique up to (unique) time-orientation-preserving isometries fixing the Cauchy surface,\footnote{
 Though rarely if ever mentioned in statements of the theorem, the isometry $\ps$ is unique (Landsman, 2022b). This can be shown by Proposition 3.62 in O'Neill (1983) or the equivalent
  argument in footnote 639 of Landsman (2021a), to the effect that  an isometry  $\ps$ is  determined at least locally (i.e.\ in a convex nbhd of $x$) by its tangent map $\ps_x'$ at some fixed $x\in M'$. Take $x\in i'(S)$. Since $\psi$ is fixed all along $ i'(S)$ by the  condition $\psi\circ i'=i$ and since it also fixes the (future-directed) normal $N_x$ to $i'(S)$ by the condition $\ps^*g=h$, it is determined locally.  Theorem 1 in Halvorson \& Manchak (2022) then applies, which is a rigidity theorem for isometries going back at least to Geroch (1969), Appendix A (as Halvorson \& Manchak acknowledge).\label{ufn}
  }   i.e.\ for any other  \mghd\ $(M',g',i')$ there exists an isometry $\ps:M'\raw M$ that preserves time orientation and satisfies  $\psi\circ i'=i$. This theorem has given rise to an ideology about \GR\  in which:
\begin{itemize}
\item All valid \emph{assumptions} about \GR\ are assumptions  about  initial data  $(S,h, K)$.
\item All valid \emph{questions} about \GR\ are  questions about ``the'' \mghd\ $(M,g,i)$ of these data. 
\end{itemize}
This \pde-based program sometimes gives a  different perspective from the Penrose--Hawking--Geroch mathematical approach to \GR\  originating in the 1960s, in which typically larger (usually maximally extended) space-times are studied. 
 In particular, in the \pde\ approach one would at first sight ask if some \mghd\ $(M,g,i)$ obtained from ``physically reasonable'' initial data satisfies strong cosmic censorship.

\noi By Theorem \ref{P79theorem} this comes down to the question if $(M,g)$ is globally hyperbolic. But this is true by construction! 
Hence  Penrose's clear and distinct formulation seems problematic in a paradoxical way:
\begin{enumerate}
\item Global hyperbolicity of the \mghd\ $(M,g,i)$ of any initial data set $(S,h, K)$ is automatic.
\item Global hyperbolicity of maximally extended (black hole) space-times is often too strong.\footnote{Maximally extended Kerr space-time is not even globally hyperbolic in the physically relevant case $0<a^2<m^2$. }
\end{enumerate}
In what they call the `evolutionary approach' to cosmic censorship,  Geroch \& Horowitz (1979), \S 5.4,  show a way out and \emph{en passant} sketch the modern \pde\ versions of both the weak and strong case. As we shall see, their version of strong cosmic censorship can actually be seen as a special case of Penrose's.

Their proposal for weak cosmic censorship is based on their earlier refinement of Penrose's notion of asymptotic flatness at null infinity (Geroch \& Horowitz, 1978). Using the notation of footnote \ref{ccfn}, assuming that 
the scaling function $\Om$ satisfies $\hat{\Dl}\,\Om=0$, as can always be achieved by a redefinition of $\Om$,
the extra condition is that \emph{the null geodesics ruling the null hypersurface $\CI^{\pm}$  are complete}.\footnote{Without this, the Penrose--Hawking  definition of a black hole as a connected component of $M\backslash I^-(\CI^+)$, with absolute event horizon $\p I^-(\CI^+)$, would be problematic. See e.g.\ footnote \ref{Heino}. At least, examples like this are avoided now.}
  \begin{quote}
\begin{small}
Consider now the following statement: for any asymptotically flat initial-data set, topologically $\R^3$, its maximal evolution [i.e.\ \mghd] is an asymptotically flat spacetime [at null infinity] (i.e.\ it even satisfies completeness of null infinity). This statement, we claim, captures a sense of cosmic censorship. That the initial-data set be topologically $\R^3$ ensures that the evolution is not singular already on $S$; that $S$ be asymptotically flat ensures that one deals with isolated systems, and in particular that any singular behavior of the evolution must be due to the system itself and not external influences. Suppose, then, that this maximal evolution were singular, say to the future of $S$. It certainly cannot be nakedly singular for the future of $S$, for $S$ must be a Cauchy surface for its evolution. Furthermore, the statement asserts that this evolution must be sufficiently large that it includes the entire asymptotic regime, so in particular asymptotic observers can live out their entire lives within this maximal evolution. What this statement means, then, is that asymptotic observers will forever be unaffected by any singular behavior of the spacetime. But this is a version of cosmic censorship. We have essentially just re-expressed our earlier statement (that the closure of $N$ have empty intersection with $\CI^+$), but now in a way that avoids the counterexample obtained by removing a point from Minkowski spacetime.\hfill (Geroch \& Horowitz, 1979, pp.\ 285--286)

In heuristic terms this means that, if we disregard exceptional initial conditions, no singularities are observed from infinity, even though  observations from infinity are allowed to continue indefinitely.\\ \mbox{}  \hfill
(Christodoulou, 1999a, p.\ A26)
\end{small}
\end{quote}
This statement of weak cosmic censorship is exemplary from a \pde\ point if view: it starts with an assumption on initial data and formulates a question about the ensuing \mghd.  However, like the definition of a black hole it depends on the concept of null infinity, which is ``prescient'' and may be seen as an undesirable idealization (cf.\ Curiel, 2019b, and footnote \ref{Cur}). Thus Christodoulou (1999a) reformulated the above statement of 
weak cosmic censorship in such a way that the idealization $\CI^+$ no longer occurs.\footnote{
 Let $(S,h,K)$ be asymptotically flat initial data for the Einstein equations (satisfying the constraints),  with \mghd\ $(M,g,i)$. Christodoulou  defines $(M,g)$  to have \emph{complete future null infinity} iff for any parameter value $s>0$ there exists a region $B_0\subset B\subset S$ such that $\p D^+(B)$, which is ruled by null geodesics, has the property that each null geodesic starting 
 in $\p J^+(B_0)\cap \p D^+(B)$ can be future extended beyond   $s$. 
 Here $D^+(B)$ is the future domain of dependence of $B$,  and each such null geodesic is supposed to have tangent vector $L=T-N$, where $T$ is the fd unit normal to $S$ in $M$ and $N$ is the outward unit normal to $\p B$ in $S$.}
 
Either way, this turns the  Penrosian version of weak cosmic censorship on its head! For whereas his version states that \emph{outgoing} signals from a black hole singularity are blocked by a (future) event horizon $H_+(\mathcal{E})$, the new version is about \emph{incoming} (null) signals: the further these are away from $H_+(\mathcal{E})$, the longer it takes them to enter $H_+(\mathcal{E})$, and in the limit (i.e., in the original formulation, at null infinity) this takes infinitely long, making future null infinity $\CI^+$ complete. 
Yet there is a  connection with Penrose:  lack of global hyperbolicity of $I^-(\CI^+)$ gives a partial Cauchy surface in $M$ a Cauchy horizon which cuts off
 $\CI^+$. This is  clear from simple examples like Schwarzschild for $m<0$,
   Reissner--Nordstr\"{o}m  at $e^2>m^2$, or Kerr at $a^2>m^2$; see e.g.\  
Dafermos \& Rodnianski (2008) and Landsman (2021a), \S 10.6.

The status of the weak cosmic censorship conjecture(s) in any reasonable form is open.\footnote{ 
For reviews see  Christodoulou (1999b, 2009), Kr\'{o}lak (2004), Joshi (2007), 
Dafermos (2012),  Bieri (2018), and   Ong (2020). N.B.\ Black hole uniqueness theorems (Chru\'{s}ciel,  Lopes Costa, \&  Heusler, 2012)  \emph{rely on} 
weak cosmic censorship!
} One especially problematic issue is the status of the genericity conditions that  define ``physically reasonable'' space-times (or initial conditions) and hence the range of applicability of the conjecture(s). For example, the physical relevance of genericity conditions typically used in the mathematical literature, which are evidently based on \pde\ techniques, has been questioned (Gundlach \& Martin-Garcia, 2007, \S 3.4).
 
 The current (\pde) version of strong cosmic censorship also goes back to Geroch \& Horowitz (1979):
   \begin{quote}
\begin{small}
Consider, for motivation, an initial-data set whose maximal evolution is extendible to the future of $S$
(\ldots) This extended spacetime cannot, by definition of the maximal evolution, have $S$ as a Cauchy surface. That is, from a point $p$ in the extension there must exist a maximally extended past-directed timelike curve which cannot be assigned a past endpoint, and which fails to meet $S$. In this rather mild sense the extended spacetime must be nakedly singular. One might therefore imagine formulating cosmic censorship as the assertion that every maximal evolution [i.e.\ \mghd] is inextendible, i.e.\ that, once the maximal evolution is completed, it is not possible to add any `extra regions' as vantage points from which observers could detect that their spacetime is singular to the future of $S$.\\ \mbox{}  \hfill (Geroch \& Horowitz, 1979, pp.\ 286--287)
 \end{small}
\end{quote}
Interestingly, they immediately refine this suggestion by giving the example of a small spacelike disk $S$ in Minkowski spacetime $\Mi$, whose \mghd\ is its domain of dependence $D(S)$ (which is a double cone), which is clearly extendible to all of $\Mi$. To avoid cases like that, they end up with the following:
   \begin{quote}
\begin{small}
\hi{Conjecture 5.2}. For $p$ any point in any extension of the maximal evolution of any non-compact initial data set $S$, $I^-(p)\cap S$ has non-compact closure.  \hfill (Geroch \& Horowitz, 1979, pp.\ 288)
 \end{small}
\end{quote}
But this refinement seems not to have survived, perhaps because the models on which the conjecture is tested already exclude trivial (counter)examples like the one mentioned. Thus we now simple have:
   \begin{quote}
\begin{small}
The appropriate notion of cosmic censorship (\ldots) is that the generic solution to Einstein's equations is globally hyperbolic,  i.e., that the maximal Cauchy development of a generic initial data set is inextendible.\hfill (Moncrief, 1981, p.\ 88)

The strong cosmic censorship conjecture says that `most' spacetimes developed as solutions of Einstein's equations from prescribed initial data cannot be extended outside of their 
maximal domains of dependence.\footnote{In \S 3 they further specify  `most' in terms of open and dense subsets in the space of initial data}\hfill (Chru\'{s}ciel,  Isenberg,  \& Moncrief, 1990, p.\ 1671)

\textbf{Conjecture 17.1} (Strong Cosmic Censorship). \emph{For generic initial data for Einstein's equations, the \mghd\ is inextendible.}   \hfill (Ringstr\"{o}m, 2009, p.\ 188)

\textbf{Conjecture 3.5} (Strong cosmic censorship). \emph{For generic asymptotically flat vacuum data sets, the maximal Cauchy development $(M,g)$ is inextendible as a suitably regular Lorentzian manifold.}\\ \mbox{}    \hfill (Dafermos, 2014b, p.\ 11)
 \end{small}
\end{quote}
If the extension satisfies the Einstein equations this is a special case of Penrose's (1979) formulation, applied to a judicious choice of space-time! To see this,  recall  the following concept (Chru\'{s}ciel, 1992):
\begin{definition}\label{Chrdef}
A   \emph{development} of initial data $(S,h,K)$ satisfying the vacuum constraints is a triple $(M,g,i)$, where $(M,g)$ is a space-time solving the vacuum Einstein equations and $i:S\raw M$ is a spacelike embedding such that $i^*g=h$ and $i(S)$ has extrinsic curvature  $K$ in $M$. Such a development is called \emph{maximal} if it has no  extension $(M',g')$ that also satisfies the vacuum Einstein equations.\footnote{ Thus the difference between a \emph{development} and a \emph{Cauchy  development}  is that in the former $i(S)$ is no longer required to be Cauchy surface in $M$, so that $(M,g)$ is not necessarily globally hyperbolic.   Chru\'{s}ciel (1992) proves \emph{existence} of maximal developments (but not \emph{uniqueness} up to isometry, as in  the globally hyperbolic case).}
\end{definition}\noi
Now apply  Penrose's strong cosmic censorship to such a maximal development, i.e.\ require it to be globally hyperbolic (cf.\ Theorem \ref{P79theorem}). The connection with inextendibility is then easily made:
\begin{proposition}\label{Chr}
A maximal development of  initial data  $(S,h,K)$ is globally hyperbolic (with Cauchy surface $i(S)$) if and only if the  \mghd\ of $(S,h,K)$
is inextendible as a solution to the vacuum Einstein equations. In that case, ``the'' maximal development  coincides with ``the'' \mghd\ (up to isomorphism). 
\end{proposition}
\emph{Proof}. The set of isometry classes $[M,g,i]$ of Cauchy developments $(M,g,i)$ of given initial data $(S,h,K)$ is partially ordered, and  the \mghd\ $[M_t,g_t,i_t]$ is its top element. Hence
if some maximal development  $(M_m,g_m,i_m)$ is globally hyperbolic, then 
$[M_m,g_m,i_m]\leq [M_t,g_t,i_t]$. On the other hand, since $(M_t,g_t,i_t)$ is a solution and $(M_m,g_m,i_m)$ is
maximal also the converse holds, so $(M_m,g_m,i_m)\cong  (M_t,g_t,i_t)$.
\enp\smallskip

 Therefore,  Penrose's conjecture applied to
the maximal development \`{a} la  Chru\'{s}ciel (1992) states that for ``generic'' initial data the \mghd\ is  inextendible \emph{as a solution to the vacuum Einstein equations}. The  tension that Penrose's  formulation may have given rise to  reflected the ambiguity about the space-time his condition of global hyperbolicity is supposed to apply to: it seems that in the 1960s and 1970s people had maximally extended solutions in mind,\footnote{Note  that although each extendible space-time has an inextendible extension (Geroch, 1970b), which may be taken as ``maximal'', this inextendible extension is not unique (except perhaps if one imposes unphysical conditions like analyticity).  }
 see e.g.\ the Penrose diagrams in Hawking \& Ellis (1973), whereas from the 1990s onward the \mghd\ in the \pde\ picture had become the space-time of choice.

However, the strong cosmic censorship conjecture as stated by all authors quoted above is  stronger than the one suggested by Proposition \ref{Chr}, in that even extensions that do not satisfy the Einstein equations are excluded. Indeed, any space-time solves these equations for \emph{some} energy-momentum tensor, however unfamiliar, and  one has no idea what kind of (strange) matter is contained in an extension.
But the Einstein equations should at least be \emph{defined}:\footnote{
Chru\'{s}ciel, Isenberg, \& Moncrief (1990) and  Chru\'{s}ciel \& Isenberg (1993) consider smooth extensions. }  to  this end one may either consider extensions in which the metric is $C^2$ (i.e.\ the borderline case where Einstein equations make sense as classical \pde s), or allow $C^0$ metrics for which the associated Christoffel symbols are locally $L^2$ (which is the least regular case in which the metric  can still be  defined as a weak solution to Einstein's equations).\footnote{See Geroch \& Traschen (1987),  Christodoulou, (2009), p.\ 9, and  Luk (2017), footnote 1. 
  Minkowski space-time  turns out to be inextendible even in $C^0$ (Sbierski, 2018), so in that case
 the validity of strong cosmic censorship is independent of the regularity of the extension. However,  for two-ended asymptotically flat data for the spherically symmetric Einstein--Max\-well-scalar field system (to which the  conjecture, so far discussed for the vacuum case, can be extended in the obvious way), the conjecture \emph{fails} in $C^0$, i.e.\ the \mghd\ is extendible with a $C^0$ metric,  but it \emph{holds} in $C^1$, in that the metric of the extension fails to be $C^1$ (Dafermos, 2003, 2005). The situation for the Kerr metric is similar   (Dafermos \& Luk, 2017).}
 
  Sensitivity to the precise formulation of genericity conditions (which from a \pde\ point of view define which \emph{initial data} are deemed ``physically reasonable'' and from a traditional \GR\ perspective do so for \emph{space-times}) is another issue;\footnote{Papers that  state such conditions include Dafermos (2003),  Ringstr\"{o}m (2009, \S 17.2.3; 2010, \S 11), and Luk \& Oh (2019a).} already the Kerr metric, in which strong cosmic censorship fails   for all parameter values (as long as $a\neq 0$ and $m\neq 0$), shows that genericity conditions are necessary. Such counterexamples
made it especially courageous of Penrose to conjecture strong cosmic censorship;
 but of course he had good arguments. His key observation, indeed one of his most prophetic insights, was first published in 1968 (i.e.\ before even  weak cosmic censorship had been formulated by him): 
 \begin{quote}
\begin{small}
There is a further difficulty confronting our observer who tries to cross [the Cauchy horizon] $H_+(\mathcal{H})$.
As he looks out at the universe he is ``leaving behind,'' he sees, in one final flash, as he crosses  $H_+(\mathcal{H})$, the entire later history of the rest of his ``old universe.'' If, for example, an unlimited amount of matter eventually falls into the star then presumably he will be confronted with an infinite density of matter along ``$H_+(\mathcal{H})$''. Even if only a finite amount of matter falls in, it may not be possible in generic situations to avoid a curvature singularity in place of  $H_+(\mathcal{H})$. This is at present an open question. But it may be, that the place to look for curvature singularities is in this region rather than (or as well as?) at the ``center.'' \hfill  (Penrose, 1968, p.\ 222)
\end{small}
\end{quote}
Here one should realize that for Penrose $H_+(\mathcal{H})$ is a (future) Cauchy horizon (relative to some partial Cauchy surface $S$) in some ``large'' (typically maximally extended) space-time, whereas from
a  \pde\ point of view it is the boundary of the \mghd\ of the corresponding initial data on $S$, if it has one; that is, if it is extendible (in a suitable regularity class). Either way,  the conjectured ``blueshift instability'' of $H_+(\mathcal{H})$ 
has  been confirmed in a large number of studies and  remains the key to strong cosmic censorship.\footnote{
Further to Simpson \& Penrose (1973), see McNamara  (1978) down to Chesler,   Narayan \&  Curiel (2020) for heuristics. Mathematically rigorous work started with 
Dafermos (2003); more recent papers may be traced back from Kehle \& Van de Moortel (2021).
 The conclusion seems to be that at least in asymptotically flat cases Cauchy horizons turn into  weak null singularities,
 which are null boundaries with $C^0$ metric  but  $\Gm_{\mu\nu}^{\rh}$ not locally  $L^2$ (Luk, 2017). At least for one-ended asymptotically flat initial data, behind such a weak null singularity there is also a strong curvature singularity at $r=0$.
  See also Luk \& Oh (2019) for the two-ended case, and  Gajic \& Luk (2017) for extremal Reissner--Nordstr\"{o}m black holes (i.e.\ $0<m=|e|$). For cosmological constant  $\lm\neq 0$ the situation changes, e.g.\ in de Sitter space ($\lm>0$) the
verdict on strong cosmic censorship depends critically on both the matter coupling and
the regularity of the extension (Dias, Reall, \& Santos, 2018), whilst in anti-de Sitter  ($\lm<0$) even weak cosmic censorship fails (Crisford \& Santos, 2017). 
The so-called \emph{cosmological case}, which is characterized by taking initial data on compact spacelike slices and indeed includes (toy) examples relevant to cosmology,  was first studied by Misner \& Taub (1969), who noted the instability of the Cauchy horizon in Taub--NUT space (this work actually predated Penrose's, see footnote \ref{Misnerfn}). See also Ryan \& Shepley (1975) and Collins \& Ellis (1979).  
} 

Failure of strong cosmic censorship is often taken to imply a failure of determinism in \GR.\footnote{See also Earman (1995),  Ringstr\"{o}m (2009),  Manchak (2017), Doboszewski (2019), and  Smeenk \& W\"{u}thrich (2021).}
The idea is that the (classical) world---including the (physical degrees of freedom of the) gravitational field itself---is governed by hyperbolic partial differential equations whose initial data should be given on a hypersurface $S$ and whose solutions should thereby be \emph{determined} on its domain of dependence $D(S)$ (Courant \& Hilbert, 1962). From Penrose's point of view, if a (typically maximally extended) space-time $(M',g')$ fails to be globally hyperbolic, then, taking some \emph{partial} Cauchy surface $S'$, neither the part
 $M' \backslash D(S')\neq \emptyset$ of space-time  behind the Cauchy horizon of $S'$ itself,
   nor things happening within space-time behind this horizon,  are determined by initial data on $S'$.
      From the \pde\ point of view, although any \mghd\ $(M,g)$ is globally hyperbolic with Cauchy surface $S\subset M$, if $(M,g)$ is extendible with extension $(M',g')$, then by definition $S$ fails to be a Cauchy surface for $M'\supset M$ and we are back to Penrose.
      
       If the extension $(M',g')$ satisfies the Einstein equations \emph{and} is globally hyperbolic, which is logically possible, then the lack of determinism is merely  apparent:  although $(M',g')$ is not determined by the initial data on $S$ \emph{originally expected to do so}, it is determined by new initial data on some Cauchy surface $S'\subset M'$, which may be accessible to physicists in $M'$. But 
    in interesting examples the \mghd\  $(M,g)$ typically has
 extensions $(M',g')$ that fail to be either 
  globally hyperbolic or unique (up to isometry), or both.\footnote{This  happens 
      e.g.\ in Taub--NUT and Gowdy cosmologies. See e.g.\ Tipler, Clarke, \& Ellis (1980), Moncrief (1981),  Chru\'{s}ciel, Isenberg, \& Moncrief (1990),  Chru\'{s}ciel \& Isenberg (1993), 
  Ringstr\"{o}m (2009, 2010), and  Doboszewski (2017). \label{Gowdy}
} In those cases the ensuing violation of determinism seems real (though hard to interpret).

In any case, although extendibility of the \mghd\ does seem to give rise to  some form of indeterminism in \GR, so that (contrapositively) determinism \emph{necessitates} strong cosmic censorship in the \pde\ sense of inextendibility of the \mghd, the latter is not \emph{sufficient} for determinism. 
The  indeterminism of someone falling into a black hole singularity is perfectly well compatible with global hyperbolicity, as  the Schwarzschild solution shows. 
More generally, in classical (mathematical) physics indeterminism may come from either a lack of 
 \emph{uniqueness}  of solutions or from a lack of
 \emph{existence} thereof; the latter includes incompleteness of solutions, i.e.\ non-existence after (or before) some finite time.
 Strong cosmic censorship  secures \emph{uniqueness} of solutions of the \emph{Einstein equations} (whose existence is given by \pde\ theory \`{a} la Choquet-Bruhat), but, as we have seen, it fails to imply
\emph{existence} (i.e.\ for all times) of solutions of the \emph{geodesic equation}  (Doboszewski, 2019). Hence at the classical level strong cosmic censorship 
  does not imply  determinism.\footnote{
In \emph{non-relativistic} mechanics bodies may disappear to infinity in finite time (Saari \& Xia, 1995), and hence, by the same (time-reversed) token, may \emph{appear} from nowhere in finite time and hence influence affairs in a way unforeseeable from any Cauchy surface.
This analogy with \GR, e.g.\ with white holes, is discussed by Earman (2007), \S 3.6.}
 But ironically, determinism might be rescued by \qm:  because of the unitarity of their time-evolution quantum objects  cannot disappear into a singularity, and hence they form standing waves. These only become sources of indeterminism  if they undergo measurement.\footnote{See Landsman (2017, 2021c, 2022a) for the final  proof of randomness of  measurement outcomes in \qm.} 

\section{Epilogue}\label{epilogue} 
 \begin{small}
\begin{quote}
Prior to the 1960s spacetime singularities were regarded as a minor embarrassment for GTR. They constituted an
embarrassment because it was thought by Einstein and others that a true singularity, a singularity in the fabric of spacetime itself, was an absurdity. But the  embarrassment was a minor one that could be swept under the rug; for the then known models of GTR containing singularities all embodied very special and physically unrealistic features. Two developments forced a major shift in attitude. First, the observation of the cosmic low temperature blackbody radiation lent credence to the notion that our universe originated in a big bang singularity.\footnote{It might be appropriate to point out that it was rather the agreement between: (i) the \emph{interpretation} of this observation as coming form a possible big bang (Jones, 2017, Part IV); (ii) Friedman-type cosmological models in \GR\ (Ellis, Maartens, \& MacCallum, 2012); (iii) the \emph{interpretation} of  redshifts of galaxies as indicating an expanding universe (Nussbaumer \& Bieri, 2009); (iv)  Gamow's earlier calculations of the abundance of light elements (Kragh, 2007, Chapter 4); (v) Hawking's singularity theorem (which was only understood by relativists), which led to the  concensus about the big bang from 1966 onwards. }
 Second, and even more importantly, a series of theorems due principally to Stephen Hawking and Roger Penrose indicated that, according to GTR, singularities cannot be relegated to the distant past because under quite general conditions they can be expected to occur both in cosmology and in the gravitational collapse of stars. Thus, singularities cannot be swept under the rug; they are, so to speak, woven into the pattern of the rug. Of course, these theorems might have been taken as turning what was initially only a minor embarrassment into a major scandal. Instead, what occurred in some quarters was a 180\degree\ reorientation in point of view: singularities were no longer relegated to obscurity; rather they were to be recognized as a central feature of the GTR, a feature which called attention to a new aspect of reality that was neglected in all previous physical theories, Newtonian and special relativistic alike.\\ \mbox{}\hfill (Earman, 1995, p.\ 65)
\end{quote}
\end{small}
Thus the way the scientific community came to believe in both black holes and a big bang in the 1960s, following 50 years of denial in the light of considerable evidence right from the start of general relativity, seems nothing short of a scientific revolution. In Kuhnian parlance, the singularity theorems of Penrose and Hawking from 1965--1966 launched  the underlying ``paradigm shift''. But here is a critical note:
 \begin{small}
\begin{quote}
 Kuhn has much of interest to say about
normal science, about the way in which a scientific community is
united by a set of practices. But what Kuhn failed to articulate is that
practices are just that---practices. They need not be, in general, statements
in which scientists (implicitly or explicitly) believe, and this for
two main reasons.
First, what unites a scientific community need not be a set of \emph{beliefs}.
Shared beliefs are much less common than shared practices. (\ldots) Second, beliefs, in themselves, cannot \emph{explain} the scientific process.
Statements lead on to statements only in the logical plane. Historically,
people must intervene to get one statement from the other. \hfill (Netz, 1999, p.\ 2)
\end{quote}
\end{small}
Much like ancient Greek mathematics (which is what Netz writes about),  singularities in \GR\ provide a fascinating interplay between \emph{practices} and \emph{beliefs}. With hindsight this interplay was almost paradoxical:
\begin{itemize}
\item Before 1965, \emph{practices} like Friedman (1922, 1924) or Lema\^{i}tre (1927, 1933) in cosmology, and Oppenheimer \& Snyder (1939) in astrophysics, which on sober judgement should have caused a paradigm shift, failed to do so, because they clashed with what the leading scientists then \emph{believed}. 
\item After 1965  the  opposite happened: at the level of  \emph{practice}, for the reasons explained in this paper (namely that it is inconclusive exactly about the  subject it should be about, namely black holes) Penrose's theorem would not have been able to launch a revolution. But  what the relevant scientific community \emph{believed} about the theorem, culminating in his 2020 Nobel Prize, carried the day. 
\end{itemize}
Finally, in the  context of the history of mathematics, Grattan-Guiness (2004)  calls attention to the historical actors' \emph{knowledge or ignorance of their ignorance}. Restricting ourselves to the two main actors:

  Did
Einstein (1918) know that he did not  know what a singular space-time was? The evidence is mixed (see \S\ref{before}): he combined egregious mistakes (in which he was not alone) with stunning foresight.

 Was Penrose (1965c) aware of the gap between his theorem and its intended application to black holes?
Here the evidence is clear: he was, and he subsequently did everything he could to overcome it. 

\addcontentsline{toc}{section}{References}
\begin{small}

\end{small}
\section*{Data availability statement}
The data for this article consist of the reasoning and the references. All references are easily found either on the internet or, in case of books,  in standard academic libraries. Readers who have any trouble finding a reference are welcome to contact the author for further help. The only exception of data not freely available is the letter from Einstein to Besso quoted in footnote 10. Those interested in seeing the entire letter should contact the Einstein archives via Hanoch Gutfreund (\verb#hanoch.gutfreund@mail.huji.ac.il# and/or Diana Buchwald (\verb#Diana_buchwald@caltech.edu#).
\end{document}